\documentclass[a4paper,11pt]{article}
\pdfoutput=1 
\usepackage{jcappub, bm, color} 
\usepackage{amssymb,amsfonts,slashed,amsthm,amsmath,graphicx, soul, empheq, cancel}
\usepackage[caption=false]{subfig}
\begin{document}

\newcommand{\vev}[1]{ \left\langle {#1} \right\rangle }
\newcommand{\bra}[1]{ \langle {#1} | }
\newcommand{\ket}[1]{ | {#1} \rangle }
\newcommand{\eV}{ \ {\rm eV} }
\newcommand{\KeV}{ \ {\rm keV} }
\newcommand{\MeV}{\  {\rm MeV} }
\newcommand{\GeV}{\  {\rm GeV} }
\newcommand{\TeV}{\  {\rm TeV} }
\newcommand{\1}{\mbox{1}\hspace{-0.25em}\mbox{l}}
\newcommand{\Red}[1]{{\color{red} {#1}}}

\newcommand{\lmk}{\left(}  
\newcommand{\rmk}{\right)}
\newcommand{\lkk}{\left[}  
\newcommand{\rkk}{\right]}
\newcommand{\lhk}{\left \{ }  
\newcommand{\rhk}{\right \} }
\newcommand{\del}{\partial}  
\newcommand{\la}{\left\langle} 
\newcommand{\ra}{\right\rangle}
\newcommand{\half}{\frac{1}{2}}

\newcommand{\bea}{\begin{array}}
\newcommand{\eea}{\end{array}}
\newcommand{\beq}{\begin{eqnarray}}
\newcommand{\eeq}{\end{eqnarray}}
\newcommand{\eq}[1]{Eq.~(\ref{#1})}

\newcommand{\dd}{\mathrm{d}}
\newcommand{\Mpl}{M_{\rm Pl}}
\newcommand{\mg}{m_{3/2}}
\newcommand{\abs}[1]{\left\vert {#1} \right\vert}
\newcommand{\mphi}{m_{\phi}}
\newcommand{\Hz}{\ {\rm Hz}}
\newcommand{\for}{\quad \text{for }}
\newcommand{\Min}{\text{Min}}
\newcommand{\Max}{\text{Max}}
\newcommand{\Kahler}{K\"{a}hler }
\newcommand{\cphi}{\varphi}
\newcommand{\Tr}{\text{Tr}}
\newcommand{\diag}{{\rm diag}}

\newcommand{\SUf}{SU(3)_{\rm f}}
\newcommand{\Upq}{U(1)_{\rm PQ}}
\newcommand{\Zpq}{Z^{\rm PQ}_3}
\newcommand{\Cpq}{C_{\rm PQ}}
\newcommand{\ubar}{u^c}
\newcommand{\dbar}{d^c}
\newcommand{\ebar}{e^c}
\newcommand{\nubar}{\nu^c}
\newcommand{\Ndw}{N_{\rm DW}}
\newcommand{\Fpq}{F_{\rm PQ}}
\newcommand{\fpq}{v_{\rm PQ}}
\newcommand{\Br}{{\rm Br}}
\newcommand{\Lag}{\mathcal{L}}
\newcommand{\Lqcd}{\Lambda_{\rm QCD}}

\newcommand{\ji}{j_{\rm inf}} 
\newcommand{\jb}{j_{B-L}} 
\newcommand{\M}{M} 
\newcommand{\im}{{\rm Im} }
\newcommand{\re}{{\rm Re} }

\def\lrf#1#2{ \left(\frac{#1}{#2}\right)}
\def\lrfp#1#2#3{ \left(\frac{#1}{#2} \right)^{#3}}
\def\lrp#1#2{\left( #1 \right)^{#2}}
\def\REF#1{Ref.~\cite{#1}}
\def\SEC#1{Sec.~\ref{#1}}
\def\FIG#1{Fig.~\ref{#1}}
\def\EQ#1{Eq.~(\ref{#1})}
\def\EQS#1{Eqs.~(\ref{#1})}
\def\TEV#1{10^{#1}{\rm\,TeV}}
\def\GEV#1{10^{#1}{\rm\,GeV}}
\def\MEV#1{10^{#1}{\rm\,MeV}}
\def\KEV#1{10^{#1}{\rm\,keV}}
\def\blue#1{\textcolor{blue}{#1}}
\def\red#1{\textcolor{blue}{#1}}

\newcommand{\eff}{\Delta N_{\rm eff}}
\newcommand{\neff}{\Delta N_{\rm eff}}
\newcommand{\cc}{\Omega_\Lambda}
\newcommand{\Mpc}{\ {\rm Mpc}}
\newcommand{\Msolar}{M_\odot}

\def\sn#1{\textcolor{red}{#1}}
\def\SN#1{\textcolor{red}{[{\bf SN:} #1]}}
\def\ft#1{\textcolor{magenta}{#1}}
\def\FT#1{\textcolor{magenta}{[{\bf FT:} #1]}}

\begin{flushright}
PNUTP-22-A11, TU-1143

\end{flushright}

\title{
Cosmological effects of Peccei-Quinn symmetry breaking on QCD axion dark matter 
}

\author{
Kwang Sik Jeong,$^{\diamondsuit}$
}
\author{
Kohei Matsukawa,$^{\spadesuit }$
}
\author{
Shota Nakagawa,$^{\spadesuit }$
}
\author{and
Fuminobu Takahashi$^{\spadesuit }$
}

\affiliation{$^\diamondsuit$ Department of Physics, Pusan National University, Busan 46241, Korea} 

\affiliation{$^\spadesuit$ Department of Physics, Tohoku University, Sendai, Miyagi 980-8578, Japan}

\abstract{
We study cosmological effects of explicit Peccei-Quinn breaking on the QCD axion dark matter. We find that
 the axion abundance  decreases or increases significantly depending on the initial position, even for a tiny Peccei-Quinn breaking that satisfies the experimental bound of the neutron electric dipole measurements. 
 If the axion first starts to oscillate around a wrong vacuum 
 and if it gets trapped there until the false vacuum disappears due to non-perturbative QCD effects, its abundance increases significantly and is independent of the decay constant $f_a$, as first pointed out in Ref.~\cite{Higaki:2016yqk}. Thus, the axion produced by the trapping mechanism can explain dark matter even when the decay constant is close to the lower limit due to stellar cooling arguments. On the other hand, if the axion starts to oscillate about a potential minimum close to the low-energy vacuum, its abundance is significantly reduced because of the adiabatic suppression mechanism. 
 This relaxes the upper limit of the axion window to large values of $f_a$.
  We also discuss how the axionic  isocurvature perturbation is affected by the Peccei-Quinn breaking term, and show that it can be suppressed in both regimes. In particular, the isocurvature bound on the inflation scale is relaxed by many orders of magnitudes for $f_a \lesssim 10^{11}\GeV$ compared to the conventional scenario.
}

\emailAdd{ksjeong@pusan.ac.kr}
\emailAdd{matsukawa@tuhep.phys.tohoku.ac.jp}
\emailAdd{shota.nakagawa.r7@dc.tohoku.ac.jp}
\emailAdd{fumi@tohoku.ac.jp}

\maketitle
\flushbottom

\section{Introduction
\label{introduction}}
The strong CP problem is one of the important unsolved problems in the Standard Model (SM), and its leading solution, the Peccei-Quinn (PQ) mechanism~\cite{Peccei:1977hh,Peccei:1977ur}, predicts the QCD axion as a very light pseudo Nambu-Goldstone boson associated with spontaneous breaking of the global U(1)$_{\rm PQ}$ symmetry~\cite{Weinberg:1977ma,Wilczek:1977pj}. The QCD axion is one of the candidates for cold dark matter (CDM) and has attracted considerable attention in recent years in terms of its cosmological evolution and direct detection experiments.
See Refs.~\cite{Jaeckel:2010ni,Ringwald:2012hr,Arias:2012az,Graham:2015ouw,Marsh:2015xka,Irastorza:2018dyq, DiLuzio:2020wdo} for reviews.

The misalignment mechanism is a simple and natural way to produce axion DM~\cite{Preskill:1982cy,Abbott:1982af,Dine:1982ah}. When the U(1)$_{\rm PQ}$ symmetry is spontaneously broken before inflation, the axion field is uniformly distributed in space and its initial value is generally deviated from the low-energy vacuum where the strong CP phase, $\bar \theta$, almost vanishes.  Also, since the axion acquires mass due to  non-perturbative effects of QCD, it is usually assumed to be nearly massless in the early universe when the temperature is above the QCD scale, $\Lambda_{\rm QCD}$, and it does not move much from its initial value until the temperature becomes comparable to $\Lambda_{\rm QCD}$. 
Then, the axion produced by the misalignment mechanism can explain the observed DM abundance, $\Omega_{\rm DM}h^2\simeq0.12$ \cite{Planck:2018vyg},  for the decay constant $f_a$ of
about $10^{12}$ GeV  and the initial misalignment angle $\theta_{\rm ini}$ of $O(1)$. This sets the upper limit of the so-called axion window,
\begin{align}
    10^8{\rm \, GeV} \lesssim f_a \lesssim 10^{12} {\rm\,GeV}.
    \label{aw}
\end{align}
Therefore, in the usual scenario, when $f_a$ is greater than $10^{12}$\,GeV, $\theta_{\rm ini}$ needs to be fine-tuned to be smaller than unity
to avoid the overproduction of the axion. One point to be noted, however, is that this  implicitly assumes an inflation scale above the QCD scale. If one considers stochastic dynamics of axion during inflation below the QCD scale, one can naturally account for the small initial value of the axion field~\cite{Graham:2018jyp,Guth:2018hsa} (see Refs.~\cite{Ho:2019ayl,Nakagawa:2020eeg,Reig:2021ipa} for the case of string axions).

The lower bound on the axion window, on the other hand, is given by the limits derived from the cooling of the neutron star~\cite{Leinson:2014ioa,Hamaguchi:2018oqw,Leinson:2019cqv,Buschmann:2021juv} and the duration of the neutrino burst from SN 1987A~\cite{Mayle:1987as,Raffelt:1987yt,Turner:1987by,Chang:2018rso,Carenza:2019pxu}. It is possible to increase the abundance of axion DM by the anharmonic effect, if the initial angle is close to $\pi$. However, at the same time, axionic isocurvature fluctuation and its non-Gaussianity are also enhanced~\cite{Lyth:1989pb,Kobayashi:2013nva}, so that the upper limit on the inflation scale becomes extremely strict.\footnote{See Ref.~\cite{Co:2018mho,Takahashi:2019pqf,Huang:2020etx} for mechanisms to set the initial position of the axion near the top of the potential.} 
While a number of direct searches for axion DM have been planned or currently ongoing in the wide range of axion masses from neV to meV, the usual misalignment mechanism does not work well to explain the right abundance of DM for $f_a$ close to the cooling bound. This is also the case when considering the decay of topological defects such as strings and domain walls.

The success of the PQ mechanism relies on high quality of the global PQ symmetry. This means that any term that breaks the PQ symmetry must be extremely suppressed. On the other hand, in quantum gravity theory, it is strongly believed that any global symmetry is explicitly broken~\cite{Kallosh:1995hi,Banks:2010zn,Witten:2017hdv,Harlow:2018jwu,Harlow:2018tng}. It is therefore important to investigate the effects of such explicit breaking of the PQ symmetry and its implications for the axion-dark matter scenario.

In this paper, we introduce a small PQ-symmetry breaking term in the axion potential and investigate its effect on the axion DM. At present, such explicit breaking of PQ symmetry is very strongly limited by neutron electric dipole moment (nEDM) measurements, but even tiny breaking can have an important impact on the axion dynamics because non-perturbative effects of QCD are suppressed in the early universe at high temperatures. We classify the possible effects on the axion dynamics and evaluate both numerically and analytically the abundance of the axion in each case, varying the relative size and  phase of the breaking term with respect to the potential coming from the QCD effects. Interestingly, we find that the axion abundance increases or decreases significantly compared to the normal scenario, depending on around which minimum of the PQ breaking term the axion first starts to oscillate. In particular, the axion abundance decreases when it starts oscillating around the minimum close to the low-energy minimum where the strong CP phase vanishes, and increases when it starts oscillating around another minimum and is trapped there for a while. Furthermore, the axion abundance when trapped is almost independent of $f_a$, indicating that the axion can explain DM even for arbitrary small $f_a$. In this case, we will see that some mild fine-tuning is needed with respect to the position of the vacuum in the PQ breaking term.
We also investigate how the axionic isocurvature fluctuation is modified in each scenario,
and show that it can be suppressed in the trapping regime. 
Finally, we briefly comment that 
the high quality of the PQ symmetry becomes closely related to the anthropic argument on 
 the axion DM abundance, once we take account of the effect of the PQ breaking on the axion dynamics.

Before proceeding, let us comment on related works in the past. In Ref.~\cite{Higaki:2016yqk}, Higaki, Kitajima, and two of the current authors (KSJ and FT)  studied the effect of the explicit PQ breaking  on the QCD axion dynamics and discussed its implications for the high quality of the PQ symmetry. They also showed that the axion abundance gets enhanced when the axion is trapped in a wrong vacuum until it starts to 
oscillate about the true vacuum due to the non-perturbative QCD effect, and that the final axion abundance is independent of $f_a$. This finding is very important, and it will also be confirmed later in this paper.\footnote{In addition, we find a mild dependence on $f_a$
when the trapping effect is not significant.}
While the results of Ref.~\cite{Higaki:2016yqk} are valid in a more general set-up, it was motivated by
the clockwork/aligned QCD axion model proposed in Ref.~\cite{Higaki:2015jag}, where the PQ breaking scale is much lower than the decay constant $f_a$, thereby suppressing the expected size of such explicit PQ breaking.\footnote{
The pre-inflationary PQ breaking scenario is assumed in Ref.~\cite{Higaki:2016yqk}. In a post-inflationary scenario, a complicated string-wall network appears when one considers the clockwork/aligned axion models, as first pointed out in Ref.~\cite{Higaki:2016jjh}. See also Refs.~\cite{Long:2018nsl,Chiang:2020aui} for subsequent developments.
}  In the present paper we do not limit ourselves to the clockwork/alignment set-up, but study the axion dynamics in the presence of the explicit PQ breaking in a more general context, for a broad range of the size and  phase of the breaking terms and various values of $f_a$. We will also study how the axionic isocurvature perturbation is affected. Various scenarios of temporarily trapping the QCD axion in a different vacuum have also been investigated in Refs.~\cite{Kawasaki:2015lpf,Nomura:2015xil,Kawasaki:2017xwt,Nakagawa:2020zjr,Kitajima:2020kig,DiLuzio:2021gos}, where the explicit PQ breaking is induced either by the Witten effect~\cite{Witten:1979ey} of hidden monopoles~\cite{Witten:1979ey,Fischler:1983sc} or $N$ mirror worlds with a $Z_N$ symmetry~\cite{Hook:2018jle}. In both cases the size of the extra PQ breaking is effectively time-dependent so that it is relevant for the axion dynamics especially at high temperatures. In  Refs.~\cite{Kawasaki:2015lpf,Nakagawa:2020zjr} it was noted that, when the PQ breaking term is sufficiently large in the early Universe, the final axion abundance is reduced due to the adiabatic suppression mechanism~\cite{Linde:1996cx, Nakayama:2011wqa} and the early oscillations. On the other hand, as mentioned above, the axion abundance can be enhanced due to the trapping.
It was pointed out in Ref.\cite{DiLuzio:2021gos} that the axion abundance gets enhanced if the axion is trapped in a wrong vacuum based on the $Z_N$ axion model, where the $Z_N$ axion mass is much lighter than the standard axion mass, and the trapping is induced by temperature-dependent effects of the mirror QCD sectors. Their trapped misalignment mechanism is analogous to the trapping phenomenon studied in Ref.~\cite{Higaki:2016yqk}. 
In the present paper we consider the explicit PQ breaking which is constant with time, as in Ref.~\cite{Higaki:2016yqk}, and will see that the axion abundance is either reduced due to the adiabatic suppression mechanism or enhanced due to the trapping, depending on around which vacuum the axion first starts to oscillate.
Also, since the axion potential in the present universe is mainly produced by non-perturbative QCD effects, the present axion mass is practically the same as the conventional one in our scenario.

The rest of this paper is organized as follows. 
In Sec.~\ref{sec:quality} we review the current experimental limits 
on the PQ symmetry breaking and identify the allowed region for the breaking parameters. In Sec.~\ref{sec:QCD-axion} we investigate 
cosmological impacts of the PQ breaking on the axion dynamics, and estimate the abundance both analytically and numerically. In Sec.~\ref{sec:isocurvature} we similarly study how the axionic isocurvature perturbation is modified.
The last section is devoted for discussion and conclusions.

\section{Experimental limits on the  PQ symmetry breaking
\label{sec:quality}}
The strong CP phase $\bar{\theta}$ is bounded by the nEDM experiments \cite{nEDM:2020crw},
\beq
|\bar{\theta}|\lesssim10^{-10},
\label{nEDM}
\eeq
where the strong CP phase is defined as the measurable quantity including the quark mass phases. 
The QCD axion is stabilized at the (nearly) CP conserving vacuum if
the PQ symmetry is explicitly broken primarily by the non-perturbative
effects of QCD. Thus, any other PQ breaking terms must be extremely suppressed for the success of the PQ mechanism, unless one of the minima of the breaking terms almost coincides with $|\bar{\theta}|=0$. In this section, we recast the above limit from the nEDM on $\bar \theta$ to that on the explicit PQ breaking term.

There is a strong argument that there is no exact continuous global symmetry in quantum gravity theory \cite{Kallosh:1995hi,Banks:2010zn,Witten:2017hdv,Harlow:2018jwu,Harlow:2018tng}, and additional PQ breaking operators, if any, could
easily spoil the PQ mechanism as a solution to the strong CP problem.
This places a non-trivial constraint on the UV theory, and it is hard to explain from low-energy perspective why the breaking of PQ symmetry other than QCD should be so small. This is known as the high quality problem of PQ symmetry.

\begin{figure}[t!]
\includegraphics[width=10cm]{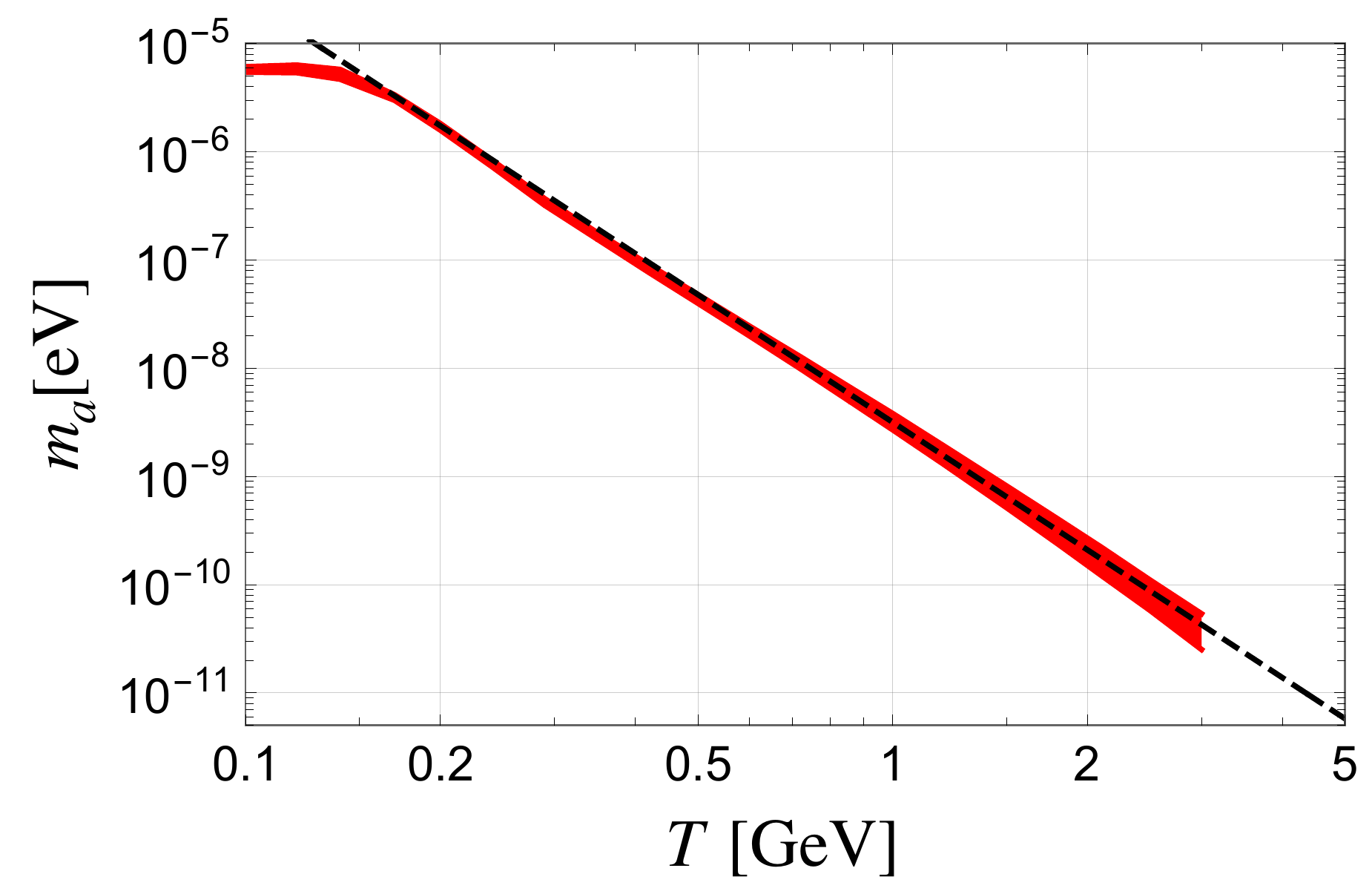}
\centering
\caption{
The temperature dependence of $m_a(T)$ obtained by the lattice result (red line)~\cite{Borsanyi:2016ksw}
for $f_a=10^{12}\GeV$.
The width of the red line represents the statistical and systematic errors. 
The black dashed line is the numerical fit obtained from
the lattice result between $T=1\GeV$ and $3\GeV$ assuming the power-law function (\ref{finiteTmass}).}
\label{latticemass}
\end{figure}

Throughout this paper we assume the pre-inflationary scenario in which  the PQ symmetry is  spontaneously broken during inflation, and
focus on the axion DM produced by the misalignment mechanism. We will briefly discuss the case of post-inflationary scenario in the last section.
The axion acquires a potential from non-perturbative QCD effects,
\beq
V_{\rm QCD}(a)
&=&m_a^2(T)f_a^2\left(1-\cos\frac{a}{f_a}\right),
\eeq
which stabilizes the axion at $\bar\theta = \langle a/f_a\rangle = 0$. Here note that the definition of the axion decay constant $f_a$ includes the domain wall number.  
In our analysis, we use the lattice results \cite{Borsanyi:2016ksw}\footnote{We use the effective degrees of freedom for energy density (for entropy density) $g_*$ ($g_{*s}$) given in Ref.~\cite{Saikawa:2018rcs} based on this lattice result.} for the temperature-dependent axion mass, $m_a(T)$, which is shown in \FIG{latticemass} in the case of $f_a=10^{12}\GeV$. The lattice result is not available at temperatures higher than $3\GeV$, but the result from $1\GeV$ to $3\GeV$ is consistent with the dilute instanton gas approximation ($m_a(T)\propto T^{-b}$, $b=4.08$) which is considered to be valid at higher temperatures. In fact, the temperature-dependent axion mass is well fitted by 
\beq
m_a(T)\simeq m_{a,0}\left(\frac{T}{\Lambda_{\rm QCD}}\right)^{-\tilde{b}},
\label{finiteTmass}
\eeq
for $T\gtrsim \Lambda_{\rm QCD}$, where we have defined $\Lambda_{\rm QCD} = 0.15\GeV$ and $\tilde{b}=3.92$. 
 Here $m_{a,0}$ is defined as the zero temperature mass which is used at $T\lesssim\Lambda_{\rm QCD}$ in our analysis.
From the chiral perturbation theory~\cite{GrillidiCortona:2015jxo,Gorghetto:2018ocs}, we have 
\beq
m_{a,0}&=&\frac{\sqrt{z}}{1+z}\frac{f_\pi m_\pi}{f_a}
\nonumber\\&\simeq&5.7\mu\hspace{-1.0mm}\eV\left(\frac{f_a}{10^{12}\GeV}\right)^{-1},
\eeq
which is consistent with the lattice result. Here $f_\pi=92.21\MeV$ is the pion decay constant, $m_\pi=135\MeV$ is the pion mass, and $z\equiv m_u/m_d$ is defined as the ratio of the up- and down-quark mass. The mass ratio is given by $z\simeq0.48$ from the average of the lattice results \cite{deDivitiis:2013xla, Basak:2015lla, Horsley:2015eaa}.
The numerical fit (\ref{finiteTmass})  is shown by the black dashed line in \FIG{latticemass}, and we will use this temperature-dependent mass to estimate the axion abundance in the next section.

Now we introduce an extra PQ breaking potential, 
\beq
V_{\cancel{{\rm PQ}}}(a)
&=&\Lambda_H^4\left[1-\cos\left(N\left(\frac{a}{f_a}-\theta_H\right)\right)\right],
\eeq
where $\Lambda_H$ denotes the potential height, $N$ is a rational number, and $\theta_H$ is a phase. We take $N>1$ so that the axion dynamics can be considerably modified by the existence of multiple minima of this potential. The size and relative phase of the PQ breaking term depend on the UV theory, and  the quality problem of the PQ symmetry  has been studied in many works \cite{Dine:1986bg, Barr:1992qq, Kamionkowski:1992mf, Kamionkowski:1992ax, Holman:1992us, Kallosh:1995hi, Carpenter:2009zs, Carpenter:2009sw}. Since we do not know which parameter region is plausible, we treat the potential height and the relative phase as free parameters.
As we will see in the next section, the axion abundance mainly depends on $\Lambda_H$ and the initial condition. 

\begin{figure}[t!]
\includegraphics[width=11cm]{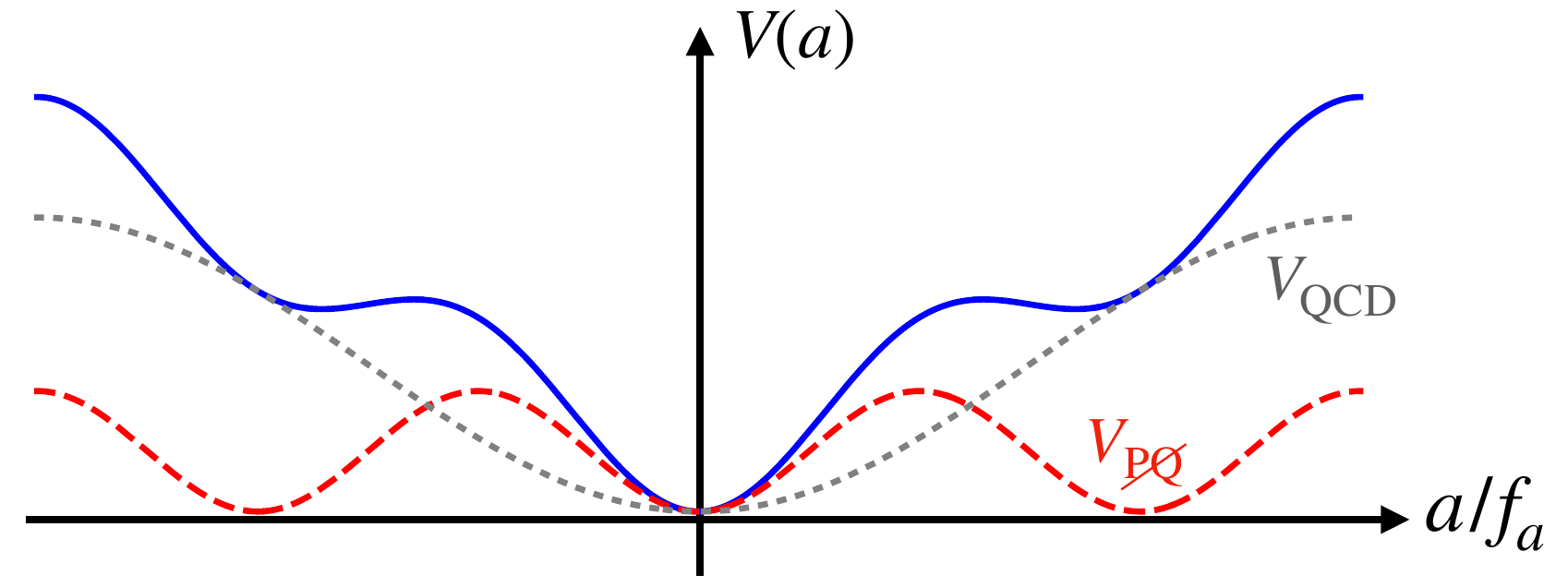}
\centering
\caption{
The schematic picture of the total potential $V(a)$ (solid blue line) with  $N=3$ and a negligibly small $\theta_H$. We also show $V_{\rm QCD}(a)$ (gray dotted), and $V_{\cancel{{\rm PQ}}}(a)$ (red dashed). The size of $V_{\cancel{{\rm PQ}}}$ is exaggerated for illustration purpose. 
}
\label{fig:shape}
\end{figure}

Now the total axion potential is given by~\cite{Higaki:2016yqk}
\beq
V(a)&=&V_{\rm QCD}(a)+V_{\cancel{{\rm PQ}}}(a)\nonumber\\
&=&m_a^2(T)f_a^2\left(1-\cos\frac{a}{f_a}\right)+\Lambda_H^4\left[1-\cos\left(N\left(\frac{a}{f_a}-\theta_H\right)\right)\right].
\label{potential}
\eeq
 The schematic picture of this potential is shown in \FIG{fig:shape}
 in the case of $N=3$ and $|\theta_H| \ll  1$. In the following we assume $\theta_H \geq 0$ without loss of generality.

Before proceeding, we comment on the potential shape of $V_{\rm QCD}$. While the dilute instanton gas approximation at $T\gg \Lambda_{\rm QCD}$ gives the cosine-type potential, the potential estimated in the chiral perturbation theory \cite{GrillidiCortona:2015jxo} at $T\ll \Lambda_{\rm QCD}$ is given by
\beq
V_{\rm ChPT}(a)=-m_\pi^2f_\pi^2\sqrt{1-\frac{4m_um_d}{(m_u+m_d)^2}\sin\left(\frac{a}{2f_a}\right)}.
\eeq
For the parameter region of our interest, the axion starts to oscillate at a temperature higher than $\Lambda_{\rm QCD}$, and the potential shape around the minimum is almost the same with the cosine potential. Thus this simplification does not affect the evaluation of the axion abundance. In fact, we have checked that this difference changes the bound from the nEDM constraint (\ref{nEDM}) only by $1\%$ or so. Therefore, we adopt the cosine-type potential with the temperature-dependent mass (\ref{finiteTmass}) in the following calculations.

\begin{figure}[t!]
\includegraphics[width=10cm]{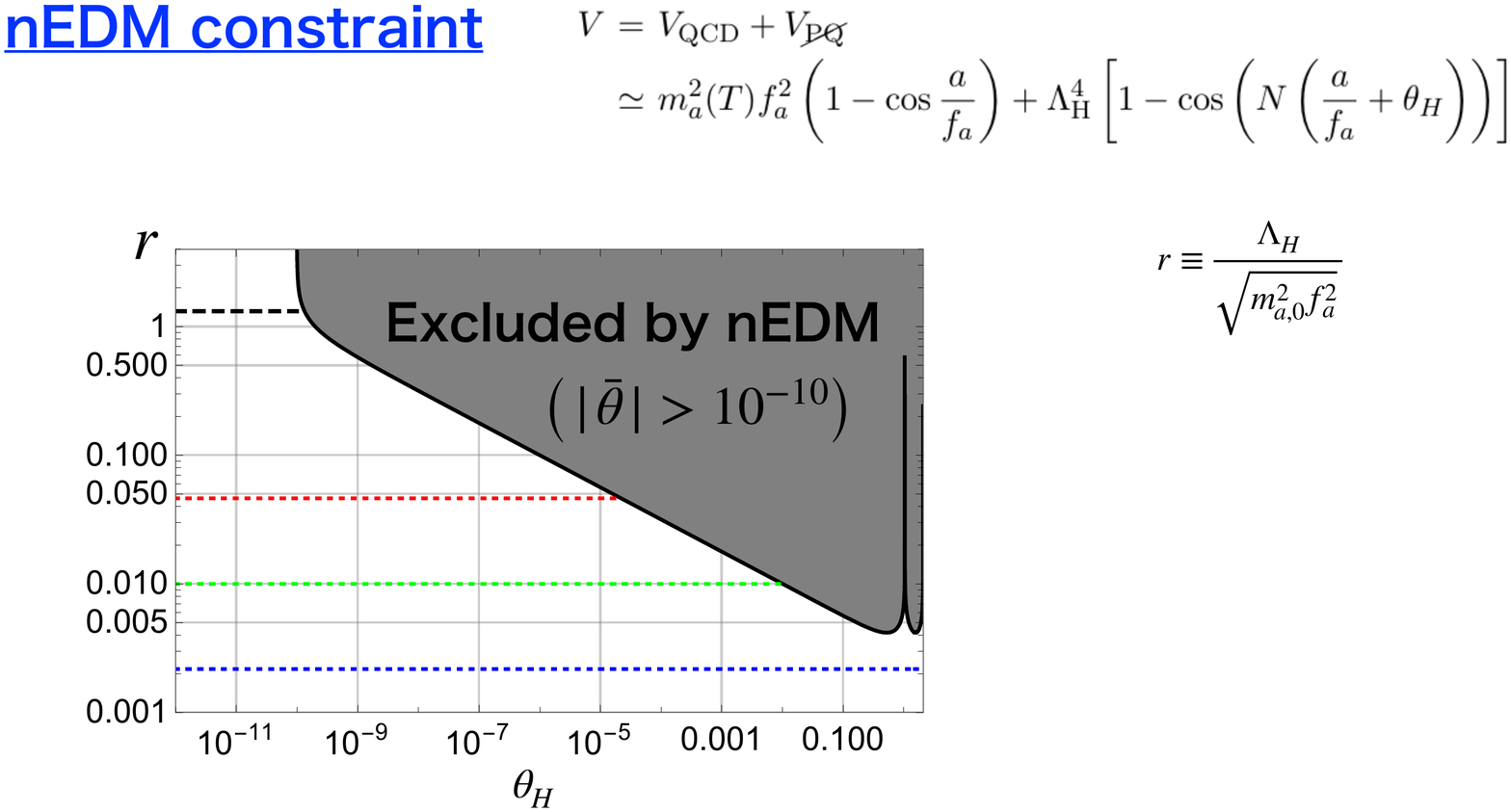}
\centering
\caption{
The nEDM constraint (\ref{constraint}) for $N=3$. The gray shaded region is excluded. The axion can fall into another minimum in the region above the black dashed line $Nr^4=1$,
in which case the contribution to $\bar\theta$ would be too large. The red, green, and blue dotted lines show the condition $T_{\rm osc}\gtrsim T^{(\rm conv)}_{\rm osc}$ for $f_a=10^{14}, 10^{12},$ and $10^{10}\GeV$, respectively (see Sec.~\ref{sec:QCD-axion}). In the region above each dotted line, the axion first starts to oscillate due to the extra PQ breaking. 
}
\label{fig:nEDM}
\end{figure}

Lastly, let us recast the nEDM bound on $\bar \theta$ to the limits on $\Lambda_H$ and $\theta_H$ which parametrize the size and phase of the extra PQ breaking. One can evaluate 
 the effective strong CP phase $\bar{\theta}$ in the presence of the extra PQ breaking term by solving $V'(a)=0$ where the prime denotes a derivative with respect to the axion field $a$. For convenience, let us define $r$ as the relative size of $V_{\cancel{{\rm PQ}}}$ with respect to $V_{\rm QCD}$, and express the limit as
 \beq
r\equiv\frac{\Lambda_H}{\sqrt{m_{a,0}f_a}}\lesssim\left|\frac{10^{-10}}{N\sin(N(10^{-10}-\theta_H))}\right|^{1/4}.
\label{constraint}
\eeq
The excluded values of $(r, \theta_H)$ are shown by the gray shaded region in \FIG{fig:nEDM} for the case of $N=3$. While the bound is absent at $|\theta_H|<10^{-10}~({\rm mod}~2\pi/N)$, the upper bound is relaxed at particular values of  $\theta_H=(2k-1)\pi/N$ with  an integer $k$
where the minimum of $V_{\cancel{{\rm PQ}}}$
is aligned with that of $V_{\rm QCD}$.\footnote{
In the current setup, we may limit  the range of $\theta_H$ as
$0\leq \theta_H < \pi/N$, although
this is not possible in the presence of another PQ breaking, in general.
} If $Nr^4\gtrsim 1$ (shown by the black dashed line), there are multiple false vacua in the low energy, and
the axion would be trapped at such a wrong vacuum if
$|\theta_{\rm ini}-\theta_H|\gtrsim \pi/N$,
where $\theta_{\rm ini} \equiv a_{\rm ini}/f_a$ is the initial angle of the axion. 
Then, it would give a too large contribution to $\bar \theta$.\footnote{If $N$ is extremely large, 
there appear multiple vacua within the nEDM bound~\cite{Higaki:2016yqk}.}

In the following sections, we will estimate the axion abundance and its isocurvature perturbation in the allowed region of $(r, \theta_H)$, and
show that they can be significantly affected by even a tiny PQ breaking that satisfies the nEDM bound.

\section{Axion abundance 
\label{sec:QCD-axion}}
The explicit breaking of the PQ symmetry can affect the dynamics of the axion, and hence its abundance. We will see that the effect is very pronounced if $r$ is large, which requires a (mild) tuning of $\theta_H$. In such a case, the abundance of axions and the isocurvature fluctuations will change significantly even when we impose the nEDM bound (\ref{constraint}).

\subsection{Dynamics of axion and analytical estimates of its abundance
\label{sec:dynamics}}
First let us study the dynamics of the QCD axion. The axion begins to move when the Hubble parameter becomes comparable to the mass or curvature of the potential. We are interested in the case in which the axion first starts to oscillate  due to the PQ breaking term well before the QCD phase transition, $T\gtrsim\Lambda_{\rm QCD}$. Otherwise the results would not differ from the conventional case, because the axion potential at the onset of oscillations would be almost identical to the conventional one.

In the conventional scenario without any extra PQ breaking,
the temperature at the onset of oscillations is given by
\beq
T^{\rm (conv)}_{\rm osc}\simeq1.1\GeV\left(\frac{g_*}{80}\right)^{-0.084}
\left(\frac{f_a}{10^{12}\GeV}\right)^{-0.17},
\eeq
where we have used $m_a(T^{(\rm conv)}_{\rm osc})=1.67H(T^{(\rm conv)}_{\rm osc})$. Throughout this paper, we assume the radiation-dominated universe until the matter-radiation equality, and the Hubble parameter is 
given by
\beq
H^2\simeq\frac{\pi^2 g_*(T)}{90} \frac{T^4}{\Mpl^2},
\eeq
where $\Mpl\equiv1/\sqrt{8\pi G}$ is  the reduced Planck mass and $g_*(T)$ is the effective relativistic degrees of freedom for energy density. 

We are interested in the case in which  the axion starts to oscillate earlier than the conventional scenario. In this case, the potential at the onset of oscillations is dominated by $V_{\cancel{{\rm PQ}}}$, and the effective mass around the minimum is approximately given by $Nr^2m_{a,0}$. Then the temperature at the onset of oscillations in this case is 
\beq
T_{\rm osc}
&\simeq&0.91\GeV\left(\frac{g_*}{80}\right)^{-1/4}\left(\frac{Nr^2}{3\times10^{-4}}\right)^{1/2}\left(\frac{f_a}{10^{12}\GeV}\right)^{-1/2},
\label{oscT}
\eeq
where we have used $Nr^2m_{a,0}=1.67H(T_{\rm osc})$. In the following, the subscript `osc' implies that the variable is evaluated  at $T=T_{\rm osc}$. We focus on the case of  $T_{\rm osc}\gtrsim T^{(\rm conv)}_{\rm osc}$, or equivalently,
\beq
Nr^2\gtrsim3.0\times10^{-4}\left(\frac{g_*}{80}\right)^{0.33}
\left(\frac{f_a}{10^{12}\GeV}\right)^{0.66}.
\label{interest}
\eeq
The lower side of the axion window (\ref{aw}) is set by
the stellar cooling arguments~\cite{Leinson:2014ioa,Hamaguchi:2018oqw,Leinson:2019cqv,Buschmann:2021juv,Mayle:1987as,Raffelt:1987yt,Turner:1987by,Chang:2018rso,Carenza:2019pxu}, and thus,
$r$ is also bounded from below as $Nr^2\gtrsim 6.7\times10^{-7}$. Note that for $f_a\gtrsim10^{11}\GeV$, some tuning of $\theta_H$ is necessary to have a sizable effect of the PQ breaking effect, while the effect can be
important for smaller $f_a$ without tuning of $\theta_H$. 
The lower bound on $r$ for different $f_a$ is shown as dotted lines in \FIG{fig:nEDM}.

The axion dynamics can be categorized into two types according to the initial position of the axion as
\begin{itemize}
     \item{Smooth shift regime} : $|\theta_{\rm ini} - \theta_H| < \pi/N$ 
    \item{Trapping regime} :  $|\theta_{\rm ini}-\theta_H| > \pi/N$.
\end{itemize}
 For integer $N$, the defining range of $\theta_{\rm ini}$ is $-\pi < \theta_{\rm ini} \leq \pi$. In the smooth shift regime, the minimum of the PQ breaking term where the axion first starts to oscillate is continuously connected to the origin where $\bar\theta$ vanishes.
 Thus, the minimum smoothly shifts to the origin  as $V_{\rm QCD}$
 becomes dominant. In this case, the final abundance of the axion is adiabatically suppressed. In the case of the trapping regime, 
 the axion first starts to oscillate in a wrong vacuum, and gets trapped
 there for a while until the false vacuum disappears when $V_{\rm QCD}$
 becomes dominant. In this case the axion abundance is enhanced. Let us estimate the axion abundance in each regime in the following.
 
 In the smooth shift regime with $|\theta_{\rm ini} - \theta_H| < \pi/N$, the axion first starts to oscillate around the minimum which is closest to $\theta = 0$.  When $V_{\rm QCD}$ becomes relevant, the minimum continuously shifts to the origin. Thus the oscillating axion adiabatically follows the temporal minimum and no extra particle production is induced during the shift of the minimum as long as the axion mass due to the PQ breaking term is much larger than the Hubble parameter. Thus,
 the axion abundance is significantly suppressed in this case. 
 This is known as the adiabatic suppression mechanism~\cite{Linde:1996cx, Nakayama:2011wqa} originally studied in a context of the Polonyi/moduli problem, which was recently applied to the axion dynamics
 in Refs.~\cite{Kawasaki:2015lpf,Kawasaki:2017xwt,Nakagawa:2020zjr}.  Then, the axion abundance is determined when the axion first starts to oscillate due to the PQ breaking term, and the initial amplitude is $|\theta_{\rm ini}-\theta_H|$. 
The ratio of the axion number density to the entropy density is given by
\beq
\left.\frac{n_a}{s}\right|_0 =\left.\frac{n_a}{s}\right|_{\rm osc}\simeq\frac{(Nr^2m_{a,0})f_a^2(\theta_{\rm ini}-\theta_H)^2/2}{s(T_{\rm osc})},
\eeq
where `0' represents the present value. Thus we obtain the axion abundance in the smooth shift regime as
\beq
\Omega^{(\rm smth)}_ah^2&=&m_{a,0}\left.\frac{n_a}{s}\right|_{\rm osc}\frac{s_0}{\rho_{\rm crit}h^{-2}}\nonumber\\
&\simeq& 5.0\times10^{-3}
F_1(\theta_{\rm ini}) \left(\frac{g_*(T_{\rm osc})}{80}\right)^{-\frac{1}{4}} (\theta_{\rm ini}-\theta_H)^2 \left(\frac{Nr^2}{3\times10^{-2}}\right)^{-\frac{1}{2}}\left(\frac{f_a}{10^{12}\GeV}\right)^{\frac{3}{2}},\nonumber\\
\label{abund1}
\eeq
where $\rho_{\rm crit}\simeq(0.003\eV)^4h^2$ is the critical density, and we assume $g_*(T_{\rm osc})\simeq g_{*s}(T_{\rm osc})$ with $g_{*s}$ defined as the effective relativistic degrees of freedom for entropy density.
Here the coefficient $F_1(\theta_{\rm ini})$ represents the contribution of the anharmonic effect for the PQ breaking term,
\beq
F_1(\theta_{\rm ini})=\left[\ln\left(\frac{e}{1-(\theta_{\rm ini}-\theta_H)^2/(\pi/N)^2}\right)\right]^{3/2},
\eeq
which is obtained following the way suggested by Refs.~\cite{Lyth:1991ub,Visinelli:2009zm} under the assumption that the PQ breaking term has no temperature dependence. 
For larger $r$, the abundance is more suppressed, because the axion starts to oscillate earlier. 

For comparison,  we give the axion abundance in the conventional scenario without the extra PQ breaking term \cite{Ballesteros:2016xej}
\footnote{The lattice result of Ref.~\cite{Borsanyi:2016ksw} was used in Ref.~\cite{Ballesteros:2016xej} to obtain the axion abundance under the harmonic approximation.},
\beq
\Omega_ah^2\simeq0.14F_2(\theta_{\rm ini})\theta_{\rm ini}^2\left(\frac{f_a}{10^{12}\GeV}\right)^{1.17},
\label{conv_abund}
\eeq
where the coefficient $F_2(\theta_{\rm ini})$ is defined as~\cite{Lyth:1991ub,Bae:2008ue,Visinelli:2009zm},
\beq
F_2(\theta_{\rm ini})=\left[\ln\left(\frac{e}{1-\theta_{\rm ini}^2/\pi^2}\right)\right]^{1.17},
\eeq
representing the contribution of the anharmonic effect for the potential from the QCD non-perturbative effect.  Here and in what follows we  assume $f_a \lesssim 10^{17} \GeV$  because the effect of the PQ breaking is negligible for very large $f_a$.
We note that (\ref{abund1}) is expected to become consistent with (\ref{conv_abund}) in the conventional case as $r$ decreases and the condition (\ref{interest}) is violated, or $T_{\rm osc}\lesssim T^{(\rm conv)}_{\rm osc}$. 

Next, let us consider the trapping regime with $|\theta_{\rm ini}-\theta_H|\gtrsim\pi/N$, where the axion is first trapped in a wrong minimum at $\theta=2\pi k/N+\theta_H$ (mod $2\pi$) with $k=1,2,... ,N-1$. 
As long as the PQ breaking term is subdominant compared to $V_{\rm QCD}$ at low temperatures, i.e. $Nr^4\lesssim 1$, all the $N-1$ local minima will disappear at a certain point, and the axion will  start to oscillate about the true minimum at $\theta = 0$. 
In this case, the axion abundance can be roughly divided into two parts: the first oscillations around the wrong vacuum, and the second oscillations around the wrong vacuum. It should be noted, however, that in practice such division involves ambiguities,
and the final abundance depends on the details of  the axion dynamics at the end of the trapping. Nevertheless, the contribution of the first oscillations is generically subdominant, since the axion abundance gets diluted by the subsequent cosmic expansion, and also the initial oscillation amplitude tends to be smaller due to the different periodicity of the two potential terms.\footnote{We will see, however, that the first oscillations 
actually give the dominant contribution to the isocurvature perturbations.}
Thus the final axion abundance is determined mainly by the difference between the disappearing local minimum and the true minimum, and it is not sensitive to the initial position $\theta_{\rm ini}$.
Such a temporal trapping of the axion in a wrong vacuum was studied in Refs.~\cite{Higaki:2016yqk,Kitajima:2020kig,DiLuzio:2021gos}, and the final axion abundance can be enhanced with respect to the conventional case.

Here we approximate the final axion abundance by the second oscillations about the true minimum. 
The axion starts to move toward the true minimum at $\theta = 0$ when the wrong vacuum disappears, i.e., $V'(a)=0$ and $V''(a)=0$.
The temperature and the oscillation amplitude satisfy the following equations,
\beq
N\tan\theta_{\rm osc2} & =&\tan[N(\theta_{\rm osc2}-\theta_H)],\\
\frac{T_{\rm osc2}}{\Lambda_{\rm QCD}}&\simeq& (Nr^4)^{-0.13}
\left[1+\left(\frac{1}{N^2}-1\right)\cos^2\theta_{\rm osc2}\right]^{0.064}.
\label{condition2}
\eeq
The subscript `osc2' implies that the variable is estimated  at this timing. Note that the first equation has $N$ solutions. One solution is the minimum closest to $\theta=0$ which corresponds to the smooth shift regime. The other solutions correspond to the wrong vacua.
If the axion is trapped in the $k$-th wrong vacuum, the oscillation amplitude will be about $\theta_{{\rm osc}2} \sim (2k-1)\pi/N$.
Note that $\theta_{\rm osc2}$ is mainly determined only by $N$ 
and $\theta_H$ but is independent of $f_a$ or $r$. 
Barring cancellation in the parenthesis in (\ref{condition2}), 
the temperature $T_{\rm osc2}$ is approximated by
\beq
T_{{\rm osc}2}\sim 0.4\GeV\left(\frac{Nr^4}{3\times10^{-4}}\right)^{-0.13}.
\eeq
Since the axion mass at $T_{\rm osc2}$ is approximately given by $m_a(T_{\rm osc2})$, we can compute the axion abundance as
\beq
\Omega^{(\rm trap)}_ah^2&=&m_{a,0}\left.\frac{n_a}{s}\right|_{{\rm osc}2}\frac{s_0}{\rho_{\rm crit}h^{-2}}\nonumber\\
&\simeq&0.25
\theta^2_{{\rm osc}2}\left(\frac{g_*(T_{{\rm osc}2})}{60}\right)^{-1}\left(\frac{Nr^4}{10^{-6}}\right)^{0.88}.
\label{abund2}
\eeq
Interestingly, the axion abundance does not depend on the decay constant because it is determined only by the potential height at $T_{\rm osc2}$. This interesting observation was first made in Ref.~\cite{Higaki:2016yqk}, and we have confirmed it here.
We note that this relation is valid when the trapping effect by the PQ breaking term is strong enough, or equivalently,
$T_{\rm osc}\gg T^{(\rm conv)}_{\rm osc}$, because we have neglected the contribution from the first oscillations. We will see in the next subsection that the abundance exhibits a mild dependence on $f_a$ as the trapping effect becomes smaller.

\subsection{Numerical calculations of the abundance
\label{sec:abundance}}
Here we present results of our numerical calculations on the axion dynamics in the presence of the extra PQ breaking. The equation of motion for the spatially homogeneous axion field is given by,
\beq
\ddot{a}+3H\dot{a}+V'(a)=0,
\label{eom}
\eeq
where the dot represents the derivative with respect to time $t$. By numerically solving this equation, we can estimate the axion abundance. More details of the numerical calculations are given in 
Appendix \ref{app:eom}.  
 
First we show in \FIG{fig:abund01m7} how the axion abundance changes with the initial angle $\theta_{\rm ini}$ for various values of the decay constant $f_a$. Here we take $N=3$, $\theta_H=10^{-7}$,
and $r=0.1(0.015)$ in the upper (lower)  panel. Note that although we have adopted a non-zero $\theta_H$, the result is not sensitive to $\theta_H$ unless it is of the order of unity, which is allowed only for $f_a \lesssim 10^{11}\GeV$. The red $(\bullet)$, orange $(\circ)$, green $(\triangle)$, dark green $(\square)$, blue $(\diamondsuit)$, and purple $(\star)$ points denote the case of $f_a=10^{15}, 10^{14}, 10^{13}, 10^{12}, 10^{11}$, and $10^{10}\GeV$, respectively,  and each colored dashed line and the black dashed line are the analytic solutions (\ref{abund1}) and (\ref{abund2}). The gray shaded region above the black dotted line represents the overproduction of the DM axion, $\Omega_a>\Omega_{\rm DM}$. The vertical gray dot-dashed lines denote the position of the local maxima of the PQ breaking term, $\theta_{\rm ini}=\pm\pi/3$, which separates the two regimes.

In the smooth shift regime,  the axion abundance increases with $|\theta_{\rm ini} - \theta_H|^2$ in good agreement with the analytical estimate (\ref{abund1}). By comparing both panels, one can also see that the axion abundance is more suppressed for larger $r$. Note that the abundance near the top of the PQ breaking term denoted by the gray dot-dashed lines is enhanced by anharmonic effect.
In the trapping regime, the axion is initially trapped at a wrong vacuum, and then starts to oscillate around the CP conserving minimum when the local minimum disappears. As a result, the abundance becomes independent of the initial position, and it can be explained well by \EQ{abund2}. For $r \lesssim 0.02$, it becomes possible to explain DM in the trapping regime (see the lower panel). 
Note also that the results for $f_a\gtrsim10^{15}\GeV$ in the upper panel or for $f_a\gtrsim10^{13}\GeV$ in the lower panel deviate from the analytical expectations. This is because the PQ breaking term  becomes relatively ineffective or $T_{\rm osc}<T^{(\rm conv)}_{\rm osc}$, which should reproduce the results for the conventional QCD axion.

\begin{figure}[t!]
\includegraphics[width=12cm]{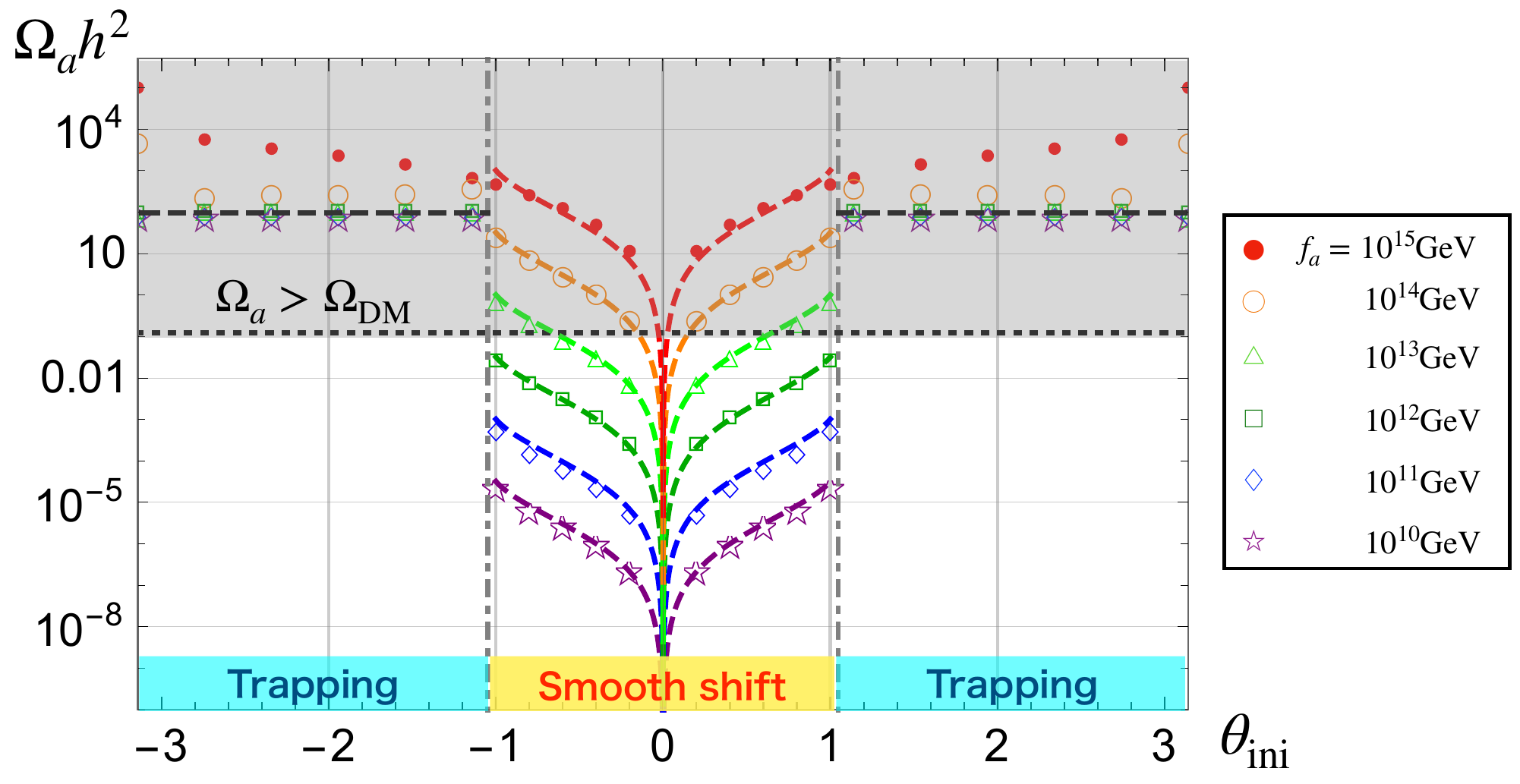}
\centering
\includegraphics[width=12cm]{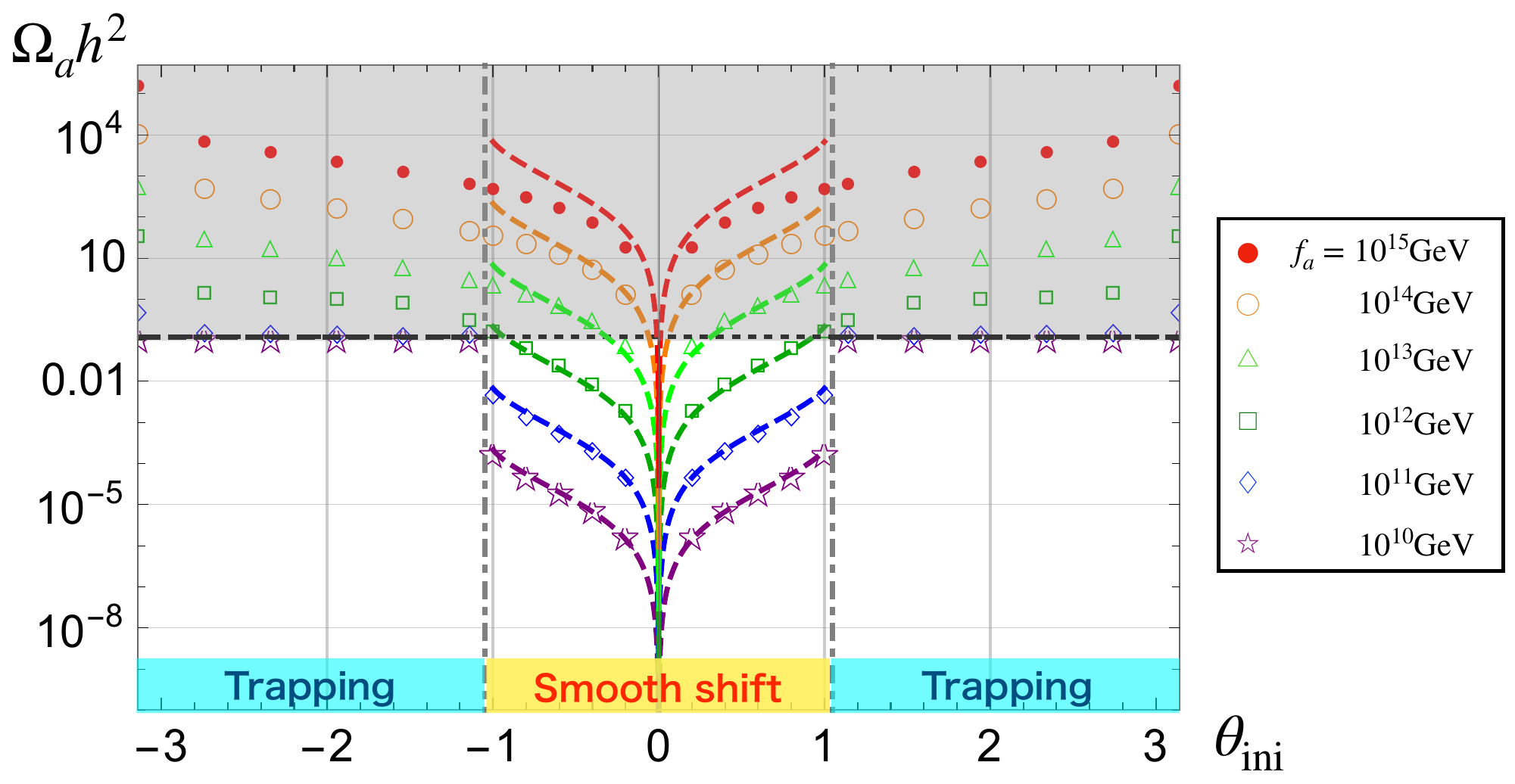}
\centering
\caption{The numerical results of the axion abundance $\Omega_ah^2$ as a function of $\theta_{\rm ini}$ for $f_a=10^{15}\GeV~({\rm red}~\bullet)$, $10^{14}\GeV~({\rm orange}~\circ)$, $10^{13}\GeV~({\rm green}~\triangle)$, $10^{12}\GeV~({\rm dark~green}~\square)$, $10^{11}\GeV~({\rm blue}~\diamondsuit)$, and $10^{10}\GeV~({\rm purple}~\star)$. In the upper panel, we take $N=3$, $r=0.1$, and $\theta_H=10^{-7}$, and in the lower panel, we take $N=3$, $r=0.015$, and $\theta_H=10^{-7}$. The colored dashed lines represent the analytical solutions (\ref{abund1}) and the black one denotes (\ref{abund2}). The gray shaded region above the black dotted line indicates the overproduction of the DM axion $\Omega_a>\Omega_{\rm DM}$. The vertical gray dot-dashed lines denote the local maxima of the PQ breaking term $|\theta_{\rm ini}-\theta_H| = \pm\pi/3$.}
\label{fig:abund01m7}
\end{figure}

\begin{figure}[t!]
\begin{minipage}[t]{8cm}
\includegraphics[width=8cm]{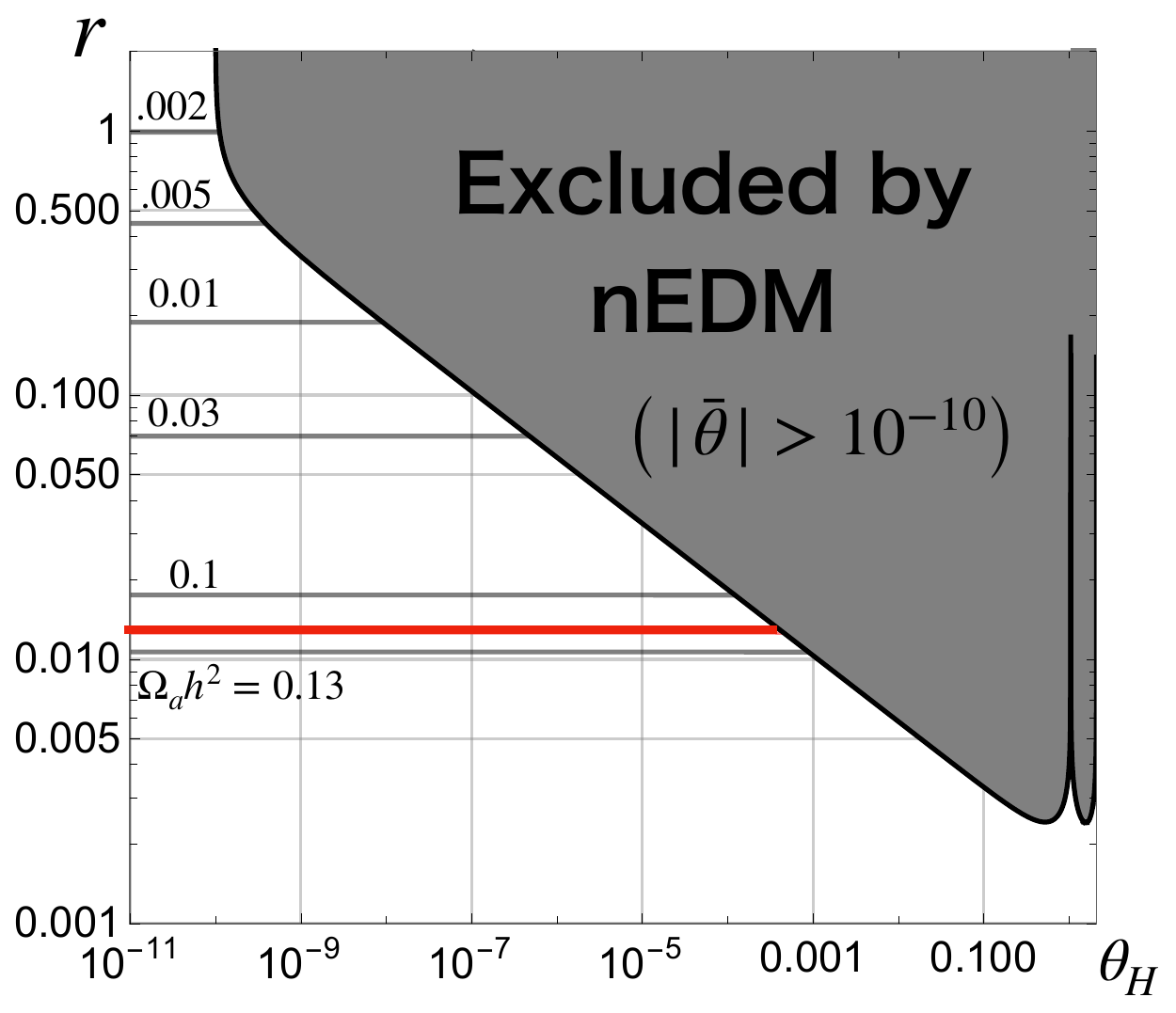}
\centering
\end{minipage}
\begin{minipage}[t]{8cm}
\includegraphics[width=8cm]{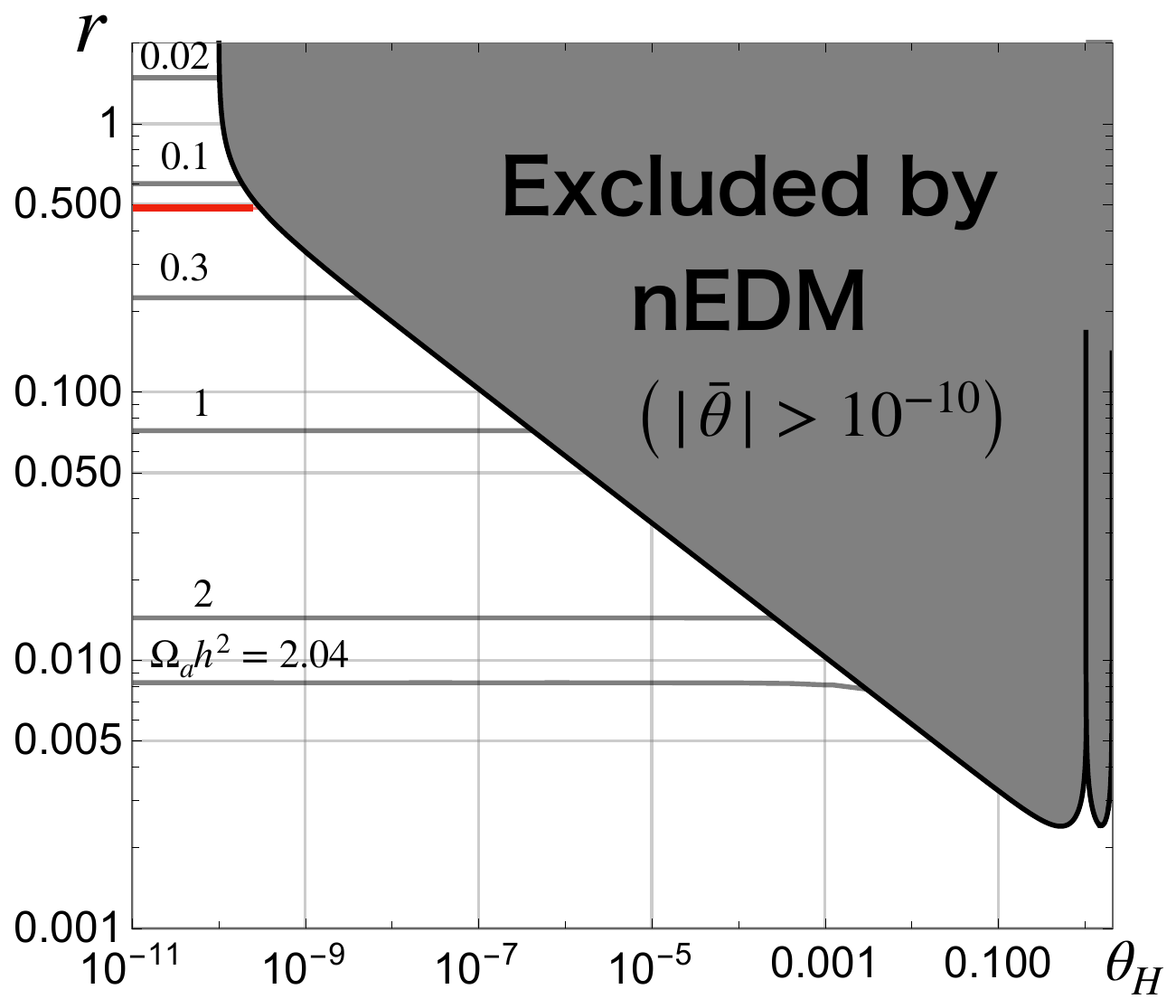}
\centering
\end{minipage}
\caption{Contours of the axion abundance $\Omega_a h^2$ in the smooth shift regime as a function of $(\theta_H, r)$ for $N=3$, and
 $\theta_{\rm ini}=1$. We set 
$f_a=10^{12}\GeV$ (left) and  $f_a=10^{13}\GeV$ (right). The red horizontal line represents $\Omega_a h^2=\Omega_{\rm DM} h^2 = 0.12$.
The axion abundance is almost constant when $r$ is sufficiently small, but gradually decreases with respect to $r$, because of the early oscillations and the adiabatic suppression.
}
\label{fig:contour}
\end{figure}

\begin{figure}[t!]
\begin{minipage}[t]{8cm}
\includegraphics[width=8cm]{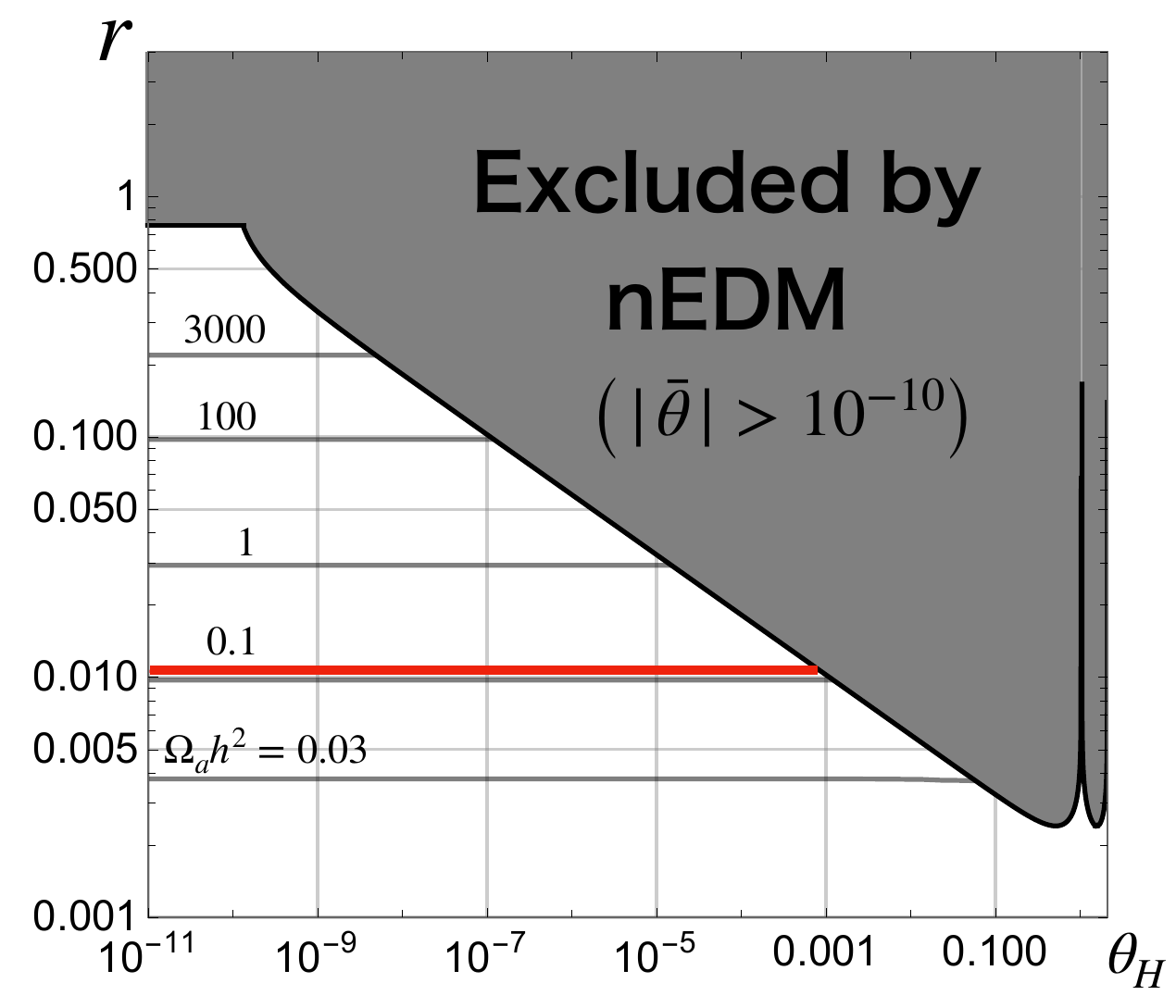}
\centering
\end{minipage}
\begin{minipage}[t]{8cm}
\includegraphics[width=8cm]{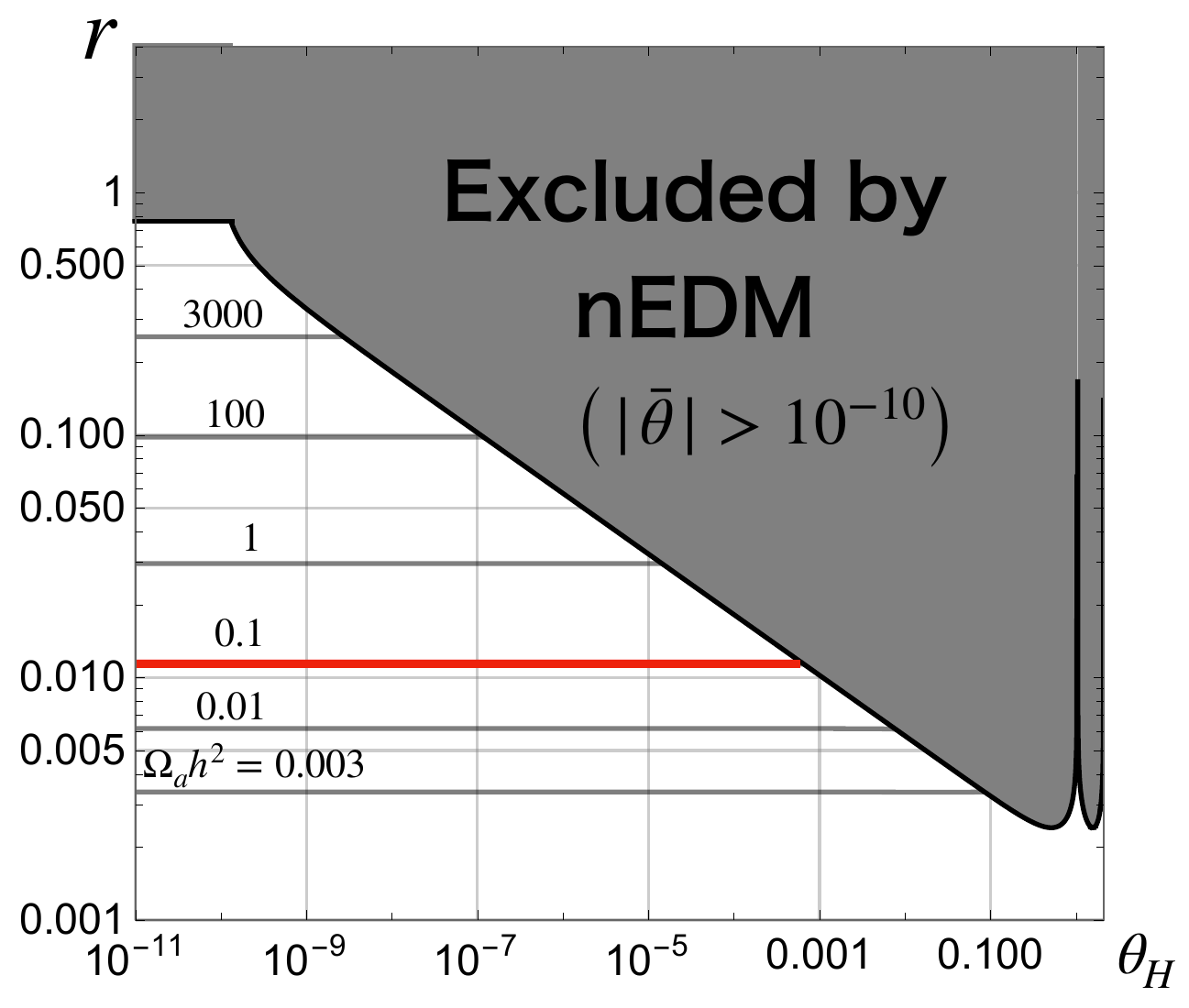}
\centering
\end{minipage}
\caption{
Contours of the axion abundance $\Omega_a h^2$ in the trapping regime as a function of $(\theta_H, r)$ for $N=3$, and
 $\theta_{\rm ini}=3/2$. We set 
$f_a=10^{11}\GeV$ (left) and  $f_a=10^{10}\GeV$ (right). The red horizontal line represents $\Omega_a h^2=\Omega_{\rm DM} h^2 = 0.12$.
The horizontal part of the gray region at small $\theta_H$ denotes the condition that the axion remains trapped in the CP violating minimum, i.e., $Nr^4\gtrsim1$.
The abundance increases with respect to $r$, and becomes independent of $f_a$}, because of the trapping effect.
\label{fig:contour2}
\end{figure}

Next we discuss how the axion abundance depends on $r$. In Figs.~\ref{fig:contour} and \ref{fig:contour2}, we
show the contour plot of the axion abundance on the $(\theta_H, r)$ plane in the smooth shift regime, and the trapping regime, respectively. In both cases we take $N=3$.
In \FIG{fig:contour} for the smooth shift regime, we set $\theta_{\rm ini}=1~ (<\pi/3)$, and $f_a=10^{12}\GeV$ (left) and $10^{13}\GeV$ (right). The gray shaded region denotes the nEDM bound.
One can see that the abundance decreases with $r$ in the region where $T^{(\rm conv)}_{\rm osc}\lesssim T_{\rm osc}$ or \EQ{interest} is satisfied. On the other hand, the abundance is almost constant for small $r$, because the axion starts to oscillate around $\theta=0$ due to $V_{\rm QCD}$, and the effect of the PQ breaking term is not important. 
 
In \FIG{fig:contour2} for the trapping regime, we set $\theta_{\rm ini}=3/2~ (>\pi/3)$, and $f_a=10^{11}\GeV$ (left) and $10^{10}\GeV$ (right). The horizontal part of the gray region represents the condition, $Nr^4\gtrsim 1$, where
the axion remains trapped a wrong vacuum, giving a too large contribution to the strong CP phase. The axion abundance increases with $r$ because the axion is trapped for a longer time, and the potential height at $T_{\rm osc2}$ becomes higher. Such $r$-dependence should be contrasted to the smooth shift regime.

\begin{figure}[t!]
\includegraphics[width=10cm]{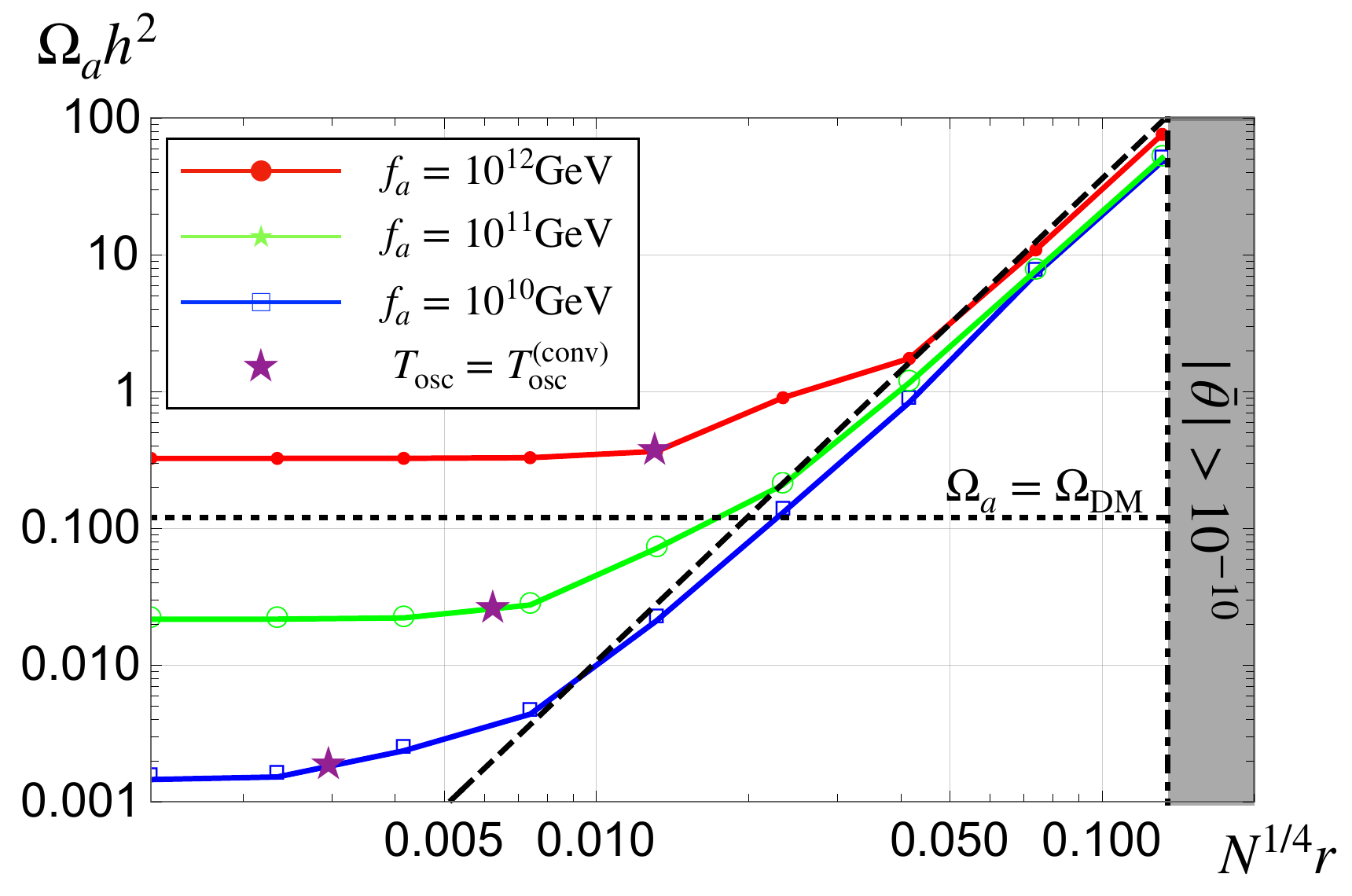}
\centering
\caption{The axion abundance in the trapping regime as a function of $N^{1/4}r$ for $f_a=10^{12}\GeV~({\rm red}~\bullet)$, $10^{11}\GeV~({\rm green}~\circ)$ and $10^{10}\GeV~({\rm blue}~\square)$. The purple stars denote the points satisfying $T_{\rm osc}=T^{(\rm conv)}_{\rm osc}$. We take $N=3$,  $\theta_H=10^{-7}$, and $\theta_{\rm ini}=3/2$. The black dashed line denotes the analytical estimate (\ref{abund2}), which is consistent with the numerical ones at $ T_{\rm osc} \gtrsim T^{(\rm conv)}_{\rm osc}$, i.e., (\ref{interest}). The black dotted line denotes the observed DM abundance. The gray shaded region represents the nEDM bound. The right DM abundance can be explained at $N^{1/4} r \sim 0.02$ for any $f_a \lesssim 10^{11}\GeV$.
}
\label{fig:largeini}
\end{figure}

Finally it is worth commenting on the dependence of $\Omega_a$ on $f_a$ in the trapping regime since the strength of the PQ breaking effect differs for different decay constants. In other words, the trapping starts later for larger $f_a$, and the effect of the PQ breaking should become relevant at larger values of $r$ for larger $f_a$. 
In \FIG{fig:largeini} we show the numerical results of $\Omega_a h^2$ as a function of $N^{1/4}r$, where we set $N=3$, $\theta_H=10^{-7}$, and $\theta_{\rm ini}=3/2~(> \pi/3)$. The red $(\bullet)$, green $(\circ)$, and blue $(\square)$ denote the abundance for $f_a=10^{12}\GeV$, $10^{11}\GeV$, and $10^{10}\GeV$, from top to bottom. The purple star on each line represents a point that satisfies $T_{\rm osc}=T^{(\rm conv)}_{\rm osc}$, and to the right of it, $T_{\rm osc}>T^{(\rm conv)}_{\rm osc}$.
The black dashed line denotes the analytic estimate (\ref{abund2}).
The horizontal black dotted  line represents the observed DM abundance, $\Omega_{\rm DM}h^2\simeq0.12$.  The gray shaded region represents the nEDM bound. On the right of each star, it is consistent with the analytic solution (\ref{abund2}), because the axion oscillation begins before the barrier at $\theta\sim\pi/N+\theta_H$ disappears. On the other hand, on the left of it, the dynamics of axion and its abundance are similar to the usual scenario.  Note that there is a small deviation among the results for different $f_a$ even in the deep trapping regime. In particular, 
the abundance is slightly larger for larger $f_a$. This is because,
when the false vacuum disappears,
it takes a bit longer for the axion to start oscillating for larger $f_a$ due to smaller hierarchy between the curvature of the potential and the Hubble parameter. This delay of the onset of the second oscillations results in a slight enhancement of the axion abundance.

\begin{figure}[t!]
\includegraphics[width=15cm]{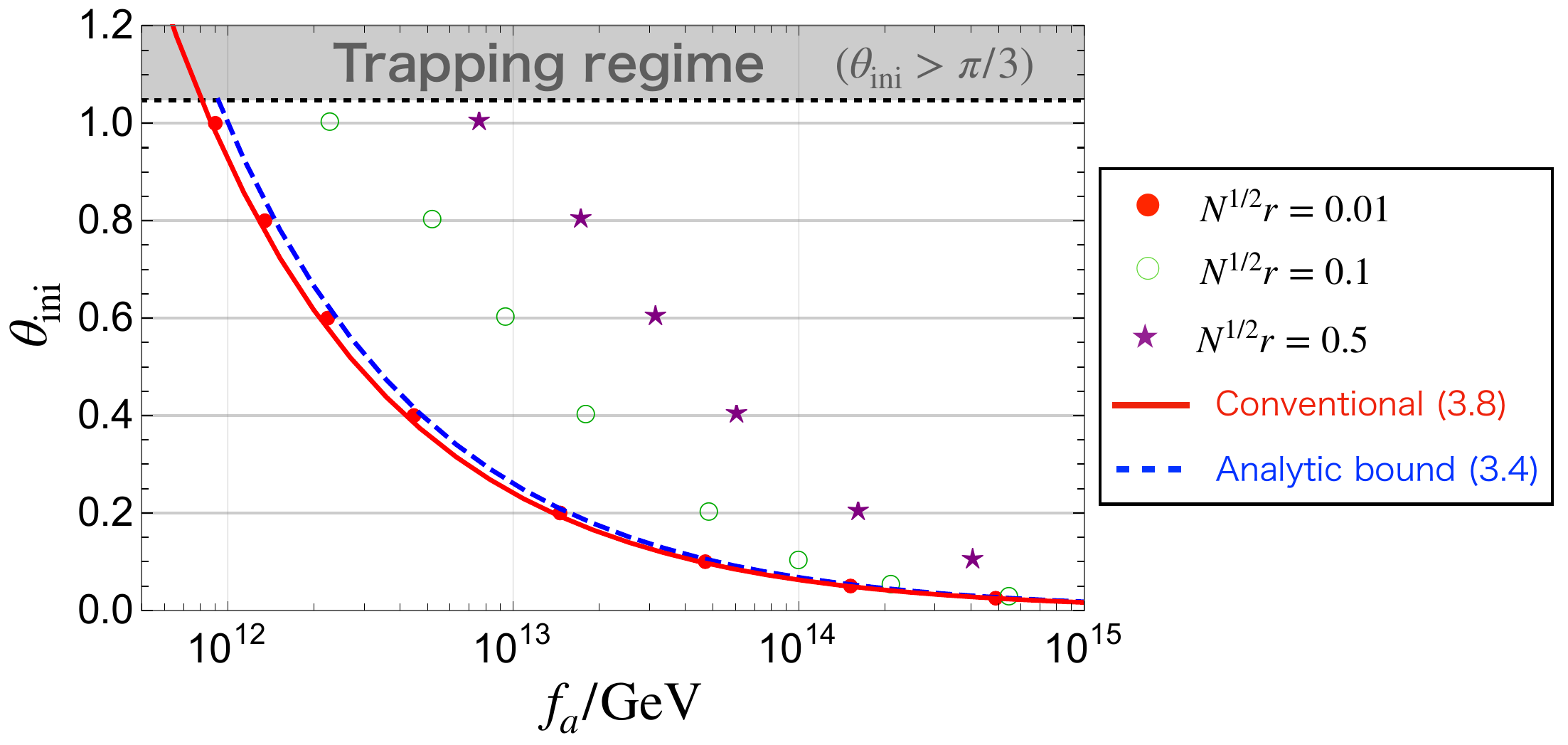}
\centering
\caption{The allowed parameter region for explaining the DM density ($\Omega_a=\Omega_{\rm DM}$) as a function of $(f_a, \theta_{\rm ini})$ in the smooth shift regime where we take $N=3$ and $\theta_H=0$. The red bullets, green circles, and purple stars denote the numerical results for $N^{1/2}r=0.01$, $0.1$, and $0.5$, respectively. The red solid and blue dashed lines represent the conventional QCD axion case (\ref{conv_abund}) and the lower bound for our interest which is analytically estimated by using (\ref{interest}) and (\ref{abund1}). The gray shaded region indicates the trapping regime.}
\label{fig:inifa}
\end{figure}

Let us briefly summarize the parameter region where it is possible to explain DM by
axion in the presence of the explicit PQ breaking. First, in the smooth shift regime, we can explain DM completely even for $f_a\gtrsim10^{12}\GeV$ due to the adiabatic suppression mechanism if $r$ is relatively large. However, such a large value of $r$ is severely constrained by nEDM, so we need some tuning of $\theta_H$, e.g. $\theta_H\lesssim10^{-6}$ for $r\simeq 0.1$. In addition, even if the tuning of $\theta_H$ is allowed, we also need a tuning of $\theta_{\rm ini}$ for $f_a\gtrsim10^{14}\GeV$. We show the allowed parameter region satisfying $\Omega_a=\Omega_{\rm DM}$ as a function of $(f_a, \theta_H)$ in \FIG{fig:inifa}, where we take $N=3$ and $\theta_H=0$. The red bullets, green circles, and purple stars denote the numerical results for $N^{1/2}r=0.01$, $0.1$, and $0.5$, respectively. The red solid line represents the case of the conventional QCD axion (\ref{conv_abund}). The blue dashed line is the bound obtained by using (\ref{interest}) and (\ref{abund1}), above which the PQ breaking term is important.
The gray shaded region indicates the trapping regime. From this figure, one can explicitly see that DM can be explained for  $f_a \gtrsim 10^{12}\GeV$ without tuning of $\theta_{\rm ini}$. At larger $f_a$,
 some tuning of $\theta_{\rm ini}$ is required, but it is still milder compared with the conventional case.  Secondly,
one can explain DM for arbitrary small $f_a$ in the trapping regime, which is very interesting.  As one can see in \FIG{fig:largeini},  if $N^{1/4} r\simeq 0.015$, the DM abundance can be totally explained  for any $f_a\lesssim10^{11}\GeV$, almost independent of $f_a$. This size of the PQ breaking requires a tuning of $|\theta_H| \lesssim 10^{-3}$. See Fig.~\ref{fig:nEDM}.

So far, we have focused on the misalignment mechanism. It may be equally interesting to consider production from decays of the topological defects in the presence of the PQ breaking in the post-inflationary scenario. We will briefly discuss this case later in \SEC{sec:conclusion}.

\section{Axionic isocurvature perturbations
\label{sec:isocurvature}}
The light axion field acquires  quantum fluctuations during inflation. If the PQ symmetry is spontaneously broken before or during inflation and is not restored afterwards, then the axionic fluctuation leaves the footprint on the  cosmic microwave background (CMB) radiation. This is the  so-called 
 isocurvature perturbation, which is tightly 
 constrained by the CMB observations. Since the size of the axion fluctuation is determined by the Hubble parameter during inflation, $H_{\rm inf}$, the isocurvature bound is usually expressed in terms of $H_{\rm inf}$ as a function of $f_a$. In this section, we briefly review the isocurvature perturbation based on 
 $\delta\mathcal{N}$ formalism \cite{Starobinsky:1985ibc,Sasaki:1995aw,Wands:2000dp,Lyth:2004gb,Kobayashi:2013nva}, and analytically and numerically estimate the axionic isocurvature perturbations in the presence of the extra PQ breaking, and compare the results with the conventional scenario.

\subsection{Analytical evaluation
\label{sec:analyiso}}
The CDM isocurvature perturbation is defined by
\beq
S(\vec{x})\equiv3(\zeta_{\rm CDM}-\zeta_{\rm radiation}),
\eeq
where $\zeta_i$ $(i={\rm CDM~or~radiation})$
denotes the curvature perturbation
on the slicing where the energy density  $\rho_i$ is uniform. 
Let us write the metric as 
$ds^2=-N_{\ell}dt^2+R^2(t)e^{2\psi}\tilde{\gamma}_{ij}(dx^i+\beta^i)(dx^j+\beta^j)$ using the  ADM decomposition \cite{Arnowitt:1962hi},
where  $N_\ell$ is the lapse function, $R(t)$  a global scale factor,
$\tilde{\gamma}_{ij}$ the spatial metric,
$\beta^i$ the shift vector, and $\psi$ is the perturbation of the expansion rate, or equivalently, the curvature perturbation. According to the  $\delta\mathcal{N}$ formalism, the difference of $\psi$ between two different time slicings is given by the difference in the e-folding number. We define the curvature perturbation on the slicing of uniform $\rho_i$  as $\zeta_i\equiv-\psi-\delta\rho_i/\bar{\rho}_i'$, where the energy density $\rho_i$ is 
decomposed into the homogeneous part and the perturbation as $\rho_i=\bar{\rho}_i+\delta\rho_i$ 
on the (total) uniform-density slicing. Here
the prime is the derivative with respect to the e-folding number, $\mathcal{N}\equiv\int Hdt$. For instance, we have approximately $\delta \rho_{\rm rad} = 0$ in the deep radiation dominated era.
 
 Each $\zeta_i$ is a conserved quantity on the superhorizon scale, as long as there is no exchange of energy between different
 energy components \cite{Lyth:2004gb}. 
 In fact, the QCD axion mass depends on temperature, and so, the axion DM and radiation do exchange the energy.
 Nevertheless, as noted in Ref.~\cite{Kobayashi:2013nva}, one can evaluate $\zeta_{\rm CDM}$ using the $\delta N$ formalism 
 after the interactions are
 turned off.\footnote{In numerical calculations, it is sufficient to follow the axion dynamics until the number-to-entropy ratio
 is fixed, since we can then compute the axion abundance when the mass becomes constant. }
 Assuming that the other cold dark matter components acquire only adiabatic fluctuations  uncorrelated with the axionic 
 fluctuation, we obtain
\beq
\zeta_{\rm CDM}+\psi=-\frac{\delta\rho_{a}+\delta\rho_{m}}{\bar{\rho}_a'+\bar{\rho}_m'}\simeq -R_a\frac{\delta\rho_{a}}{\bar{\rho}_{a}'},
\eeq
with $R_a\equiv\Omega_a/\Omega_{\rm DM}$. Here $\bar{\rho}_{a(m)}$ and $\delta\rho_{a (m)}$ denote the homogeneous part and the fluctuation of the axion (the other CDM) energy density, respectively. Note that we have used $\delta \rho_m = 0$ in the second equality. This is because we have approximately $\delta\rho_{\rm rad} = \delta\rho_{m} =0$ in the deep radiation dominated era.

We can evaluate $S(\vec{x})$  after the final axion abundance is fixed.
We obtain the CDM isocurvature perturbation using the fluctuation of the e-folding number $\delta\mathcal{N}$ \cite{Kawasaki:2008sn,Langlois:2008vk,Kawasaki:2008pa},
\beq
S(\vec{x})\simeq-3R_a\frac{\delta\rho_{a}}{\bar{\rho}_{a}'}\simeq3R_a\delta\mathcal{N}.
\eeq
Note that $\delta\mathcal{N}$ is defined as the fluctuation of the number of e-folds from the initial flat slicing at the horizon exit of the CMB scales to the uniform-$\rho_a$ slicing after the axion abundance gets fixed. We expand $\delta\mathcal{N}$ in terms of the axion fluctuation as
\footnote{We note that this expansion is not always justified. In particular, if the  e-folding number is extremely sensitive to the initial position, such an expansion could break down. 
This might be the case of the deep trapping regime because even a tiny difference in the initial condition could result in a large difference in the oscillation phase after many oscillations. We have confirmed the validity of this expansion 
in the parameter region studied in this paper.
}
\beq
\delta\mathcal{N}\simeq\frac{\del\mathcal{N}}{\del a_*}\delta a_* + \frac{1}{2}\frac{\del^2\mathcal{N}}{\del a_*^2}(\delta a_*^2-\langle\del a_*^2\rangle)+\cdot\cdot\cdot,
\label{expand}
\eeq
where $\delta a_* = H_{\rm inf}/2\pi$ represents the initial fluctuation of the axion field at the horizon exit, and $a_*$ is identified with the initial position $a_{\rm ini}$ in the previous section. The isocurvature power spectrum is defined as
\beq
\langle \mathcal{S}(\vec{k}_1)\mathcal{S}(\vec{k}_2)\rangle\equiv(2\pi)^3\mathcal{P}_S(\vec{k}_1)\delta^{(3)}(\vec{k}_1+\vec{k}_2),
\eeq
where $\mathcal{S}(\vec{k})$ denotes the Fourier component of the isocurvature perturbation $S(\vec{x})$. Taking account of the leading term of (\ref{expand}) and using the power spectrum of $\delta a_*$, $\mathcal{P}_{\delta a_*}(\vec{k})=H_{\rm inf}^2/2k^3$, we obtain the dimensionless power spectrum,
\beq
\Delta_S^2\equiv\frac{k^3}{2\pi^2}\mathcal{P}_S(\vec{k})
\simeq\left(3R_a\frac{\del\mathcal{N}}{\del \theta_{\rm ini}}\frac{H_{\rm inf}}{2\pi f_a}\right)^2
\equiv
(R_a\Delta_a)^2.
\label{formula}
\eeq
where we have defined $\Delta_a$ for later use. This formula is a useful form to estimate the isocurvature perturbation numerically. Noting that the e-folding number is measured from the initial flat slicing to the final slicing of $\delta\rho_a=0$, we can obtain its derivative by taking the difference of the e-folding numbers for slightly different  initial positions. Numerically, it is sufficient to follow each e-folding number until  the axion number density becomes equal to a fixed value after $n_a/s$ becomes constant with time. Specifically, we have the relation $n_{\rm I}(T_E)=n_{\rm II}(T_E+\Delta T)$, where $T_E$ is an arbitrary temperature (lower than $T_{\rm osc}$ or $T_{\rm osc2}$) on the final slicing, $n_{\rm I}$ is the number density for the initial position $\theta_{\rm ini}$, and $n_{\rm II}$ is the number density for $\theta_{\rm ini}+\Delta\theta$. After the (final) coherent oscillation started, $n/s$ becomes constant with time, and so we get
\beq
\frac{s(T_E+\Delta T_E)}{s(T_E)}=\frac{n_{\rm I}/s}{n_{\rm II}/s}.
\eeq
Thus the difference of the e-folding number is written by
\beq
\Delta\mathcal{N}=-\frac{1}{3}\ln\left[\frac{s(T_E+\Delta T)}{s(T_E)}\right]=-\frac{1}{3}\ln\left[\frac{n_{\rm I}/s}{n_{\rm II}/s}\right],
\eeq
which  is independent of the temperature $T_E$ but depends only on $\Delta\theta$ as we expected. Thus we can evaluate the isocurvature perturbation by calculating the ratio of $n/s$ for slightly different initial conditions.

Using (\ref{formula}), it is also possible to obtain the analytical formula \cite{Kobayashi:2013nva},
\beq
\Delta_S^2\simeq\left(R_a\frac{\del\ln\Omega_a}{\del\theta_{\rm ini}}\frac{H_{\rm inf}}{2\pi f_a}\right)^2.
\label{analyticiso}
\eeq
Although it is straightforward to estimate the isocurvature perturbation from this formula in the smooth shift regime, we need somewhat careful estimate in the trapping regime because the axion abundance is determined mainly by the amplitude $\theta_{\rm osc2}$ at $T_{\rm osc2}$ and thus is not apparently dependent on $\theta_{\rm ini}$. Noting that the axion fluctuation $\delta a$ is diluted in proportion to $R^{-3/2}$ when the axion oscillates around the quadratic potential, we obtain the fluctuation at $T_{\rm osc2}$ as, $\delta a_{\rm osc2}\simeq(T_{\rm osc2}/T_{\rm osc})^{3/2}\delta a_*$. Thus, in the trapping regime, the isocurvature perturbation is given by
\beq
\Delta_S^2\simeq\left(R_a\frac{\del\ln\Omega_a}{\del\theta_{\rm osc2}}\left(\frac{T_{\rm osc2}}{T_{\rm osc}}\right)^{\frac{3}{2}}\frac{H_{\rm inf}}{2\pi f_a}\right)^2.
\label{analyticiso2}
\eeq
The suppression factor can be written as
\beq
\left(\frac{T_{\rm osc2}}{T_{\rm osc}}\right)^{\frac{3}{2}}
&=&\left(\frac{(1.67)^2\pi^2g_*(T_{\rm osc})\Lambda_{\rm QCD}^4}{90N\Mpl^2m_{a,0}^2}\right)^{\frac{3}{8}}\cdot(N^{\frac{1}{4}}r)^{-\frac{2}{\tilde{b}}-1}\nonumber\\
&\simeq&1.5\times10^{-2}N^{-\frac{3}{8}}\left(\frac{Nr^4}{3\times10^{-4}}\right)^{-0.57}
\left(\frac{f_a}{10^{12}\GeV}\right)^{\frac{3}{4}}.
\label{SF}
\eeq
The suppression becomes stronger for larger $r$ and smaller $f_a$. Note that the above estimate does not take account of 
 the anharmonic effect of the oscillations around the false vacuum, nor
the axion dynamics when the false vacuum disappears and the trapping ends. In fact it is not possible to completely separate the trapping regime from the subsequent oscillations, and the above analytical estimate should be considered as a rough order evaluation.
Further refinements require taking account of the effects of the velocity of the axion field and the evolution of the axion fluctuation in a time-dependent potential that deviates significantly from the quadratic one toward the end of the trapping, which is beyond the scope of this paper. As we shall see shortly, however, the above analytical estimate gives an overall good fit to the numerical results, and the dependence on $\theta_{\rm ini}$ is relatively mild.

The recent Planck data constrains the scale-invariant and uncorrelated isocurvature perturbation as \cite{Planck:2018jri}:
\beq
\beta_{\rm iso}(k_0)<0.038~~(95\% {\rm CL}),
\eeq
where $k_0=0.05\Mpc^{-1}$ and $\beta_{\rm iso}$ is defined as the ratio between the power spectrum of the adiabatic perturbation and the isocurvature one, $\beta_{\rm iso}\equiv\mathcal{P}_S/\mathcal{P}_\zeta$. Thus the current upper bound on the isocurvature power spectrum reads
\beq
\Delta_S^2<8.3\times10^{-11},
\label{isoconstraint}
\eeq
which severely constrains the inflation scale $H_{\rm inf}$.

\subsection{Numerical results
\label{sec:numeriso}}
\begin{figure}[t!]
\begin{minipage}[t]{8cm}
\includegraphics[width=8cm]{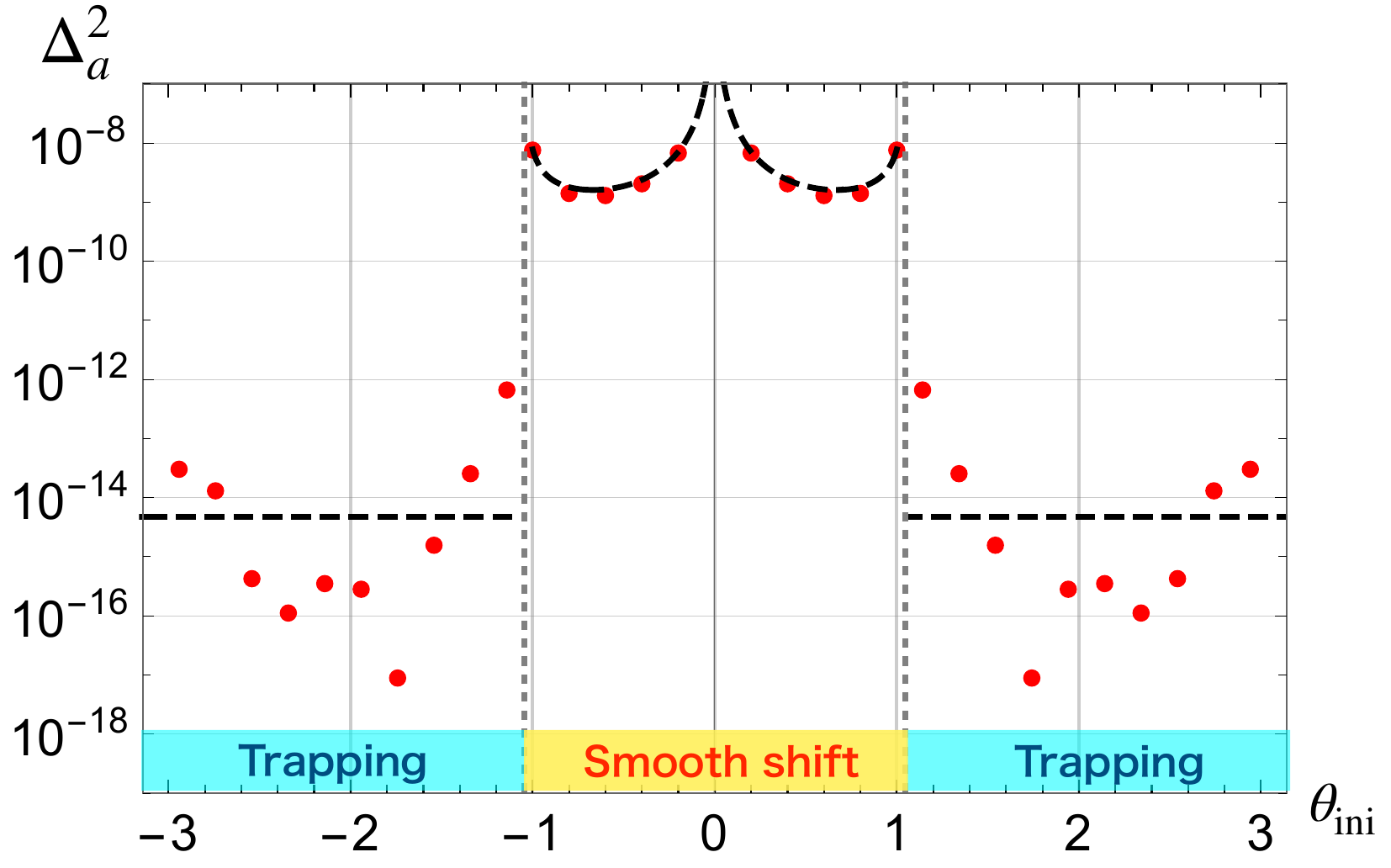}
\centering
\end{minipage}
\begin{minipage}[t]{8cm}
\includegraphics[width=8cm]{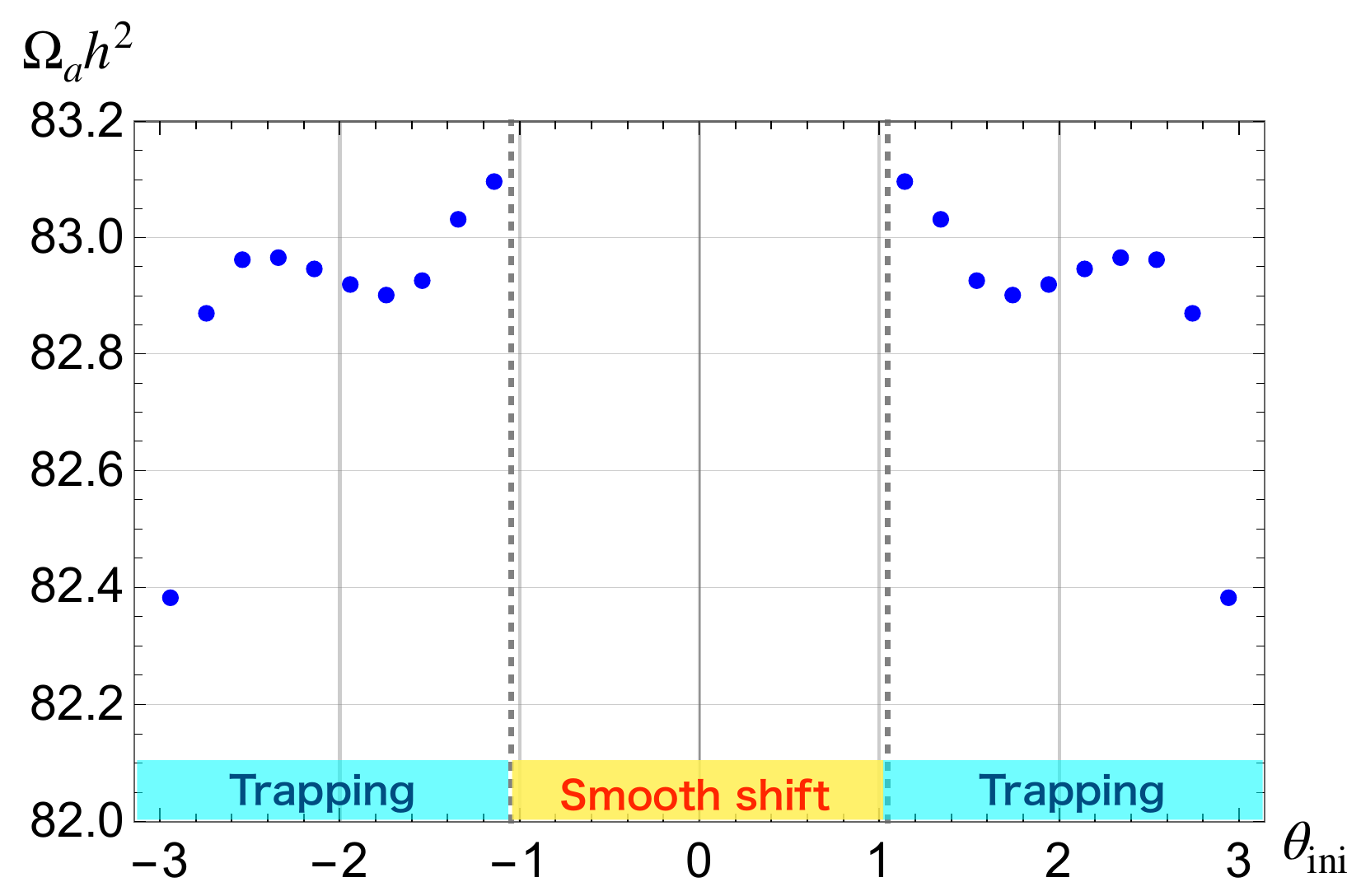}
\centering
\end{minipage}
\caption{The isocurvature power spectrum (left) and the axion abundance (right) as a function of the initial position. We set $N=3$, $r=0.1$, $\theta_H=10^{-7}$, $f_a=10^{12}\GeV$, and $H_{\rm inf}=5\times10^{7}\GeV$. The red (blue) bullets denote the numerical results of the isocurvature power spectrum (abundance), and the black dashed line is the analytical formula (\ref{analyticiso}) and (\ref{analyticiso2}). The gray dotted lines represent the maxima of the PQ breaking term.}
\label{fig:result1}
\end{figure}

Here we present our numerical results of the axionic isocurvature power spectrum. In \FIG{fig:result1} we show the results of the isocurvature power spectrum (left) and the axion abundance in the trapping regime (right) as a function of $\theta_{\rm ini}$. We set $N=3$, $r=0.1$, $\theta_H=10^{-7}$, $f_a=10^{12}\GeV$, and $H_{\rm inf}=5\times10^{7}\GeV$. 
We note again that the result is not sensitive to $\theta_H$ unless it is of the order of unity. The red and blue bullets denote the numerical values of $\Delta_a^2$ and $\Omega_a h^2$. The black dashed lines are the analytical results, (\ref{analyticiso}) and (\ref{analyticiso2}). In the smooth shift regime, the numerical results are consistent with the analytical one, 
and 
the anharmonic effect starts to be effective and enhances the isocurvature perturbation
at $\theta_{\rm ini}\gtrsim \pi/6$.
In the trapping regime, the isocurvature perturbation is significantly suppressed because the axion fluctuation is damped due to the cosmic expansion from the onset of the first oscillations until the secondary oscillations. 
Note that the time derivative of the axion field at $T\sim T_{\rm osc2}$ makes the $\theta_{\rm ini}$-dependence complicated in the trapping regime. When the potential barrier disappears at $T\sim T_{\rm osc2}$, the axion generically has a small velocity which depends on the evolution during many oscillations in the false vacuum as well as the anharmonicity of the PQ breaking term. If the axion has a velocity in the opposite direction with the shift of the potential minimum, then its abundance is slightly enhanced because 
the onset of oscillations is delayed. In the opposite case, the abundance is suppressed for similar reason. 
As a result, one can see complicated wiggles in the behavior of the axion abundance as well as the isocurvature power spectrum. Such complicated dependence is not taken into account in our  analytic estimate. Turning to the dependence of the axion abundance on $\theta_{\rm ini}$, we find that it has a local minimum and maximum at $|\theta_{\rm ini}| \approx 1.7$ and $2.4$. At these points, $\partial \Omega_a/\partial \theta_*$ vanishes, which means that the isocurvature fluctuations disappear in the linear approximation. Indeed, it can be seen that the isocurvature fluctuations are very small around the corresponding initial values.

\begin{figure}[t!]
\includegraphics[width=10cm]{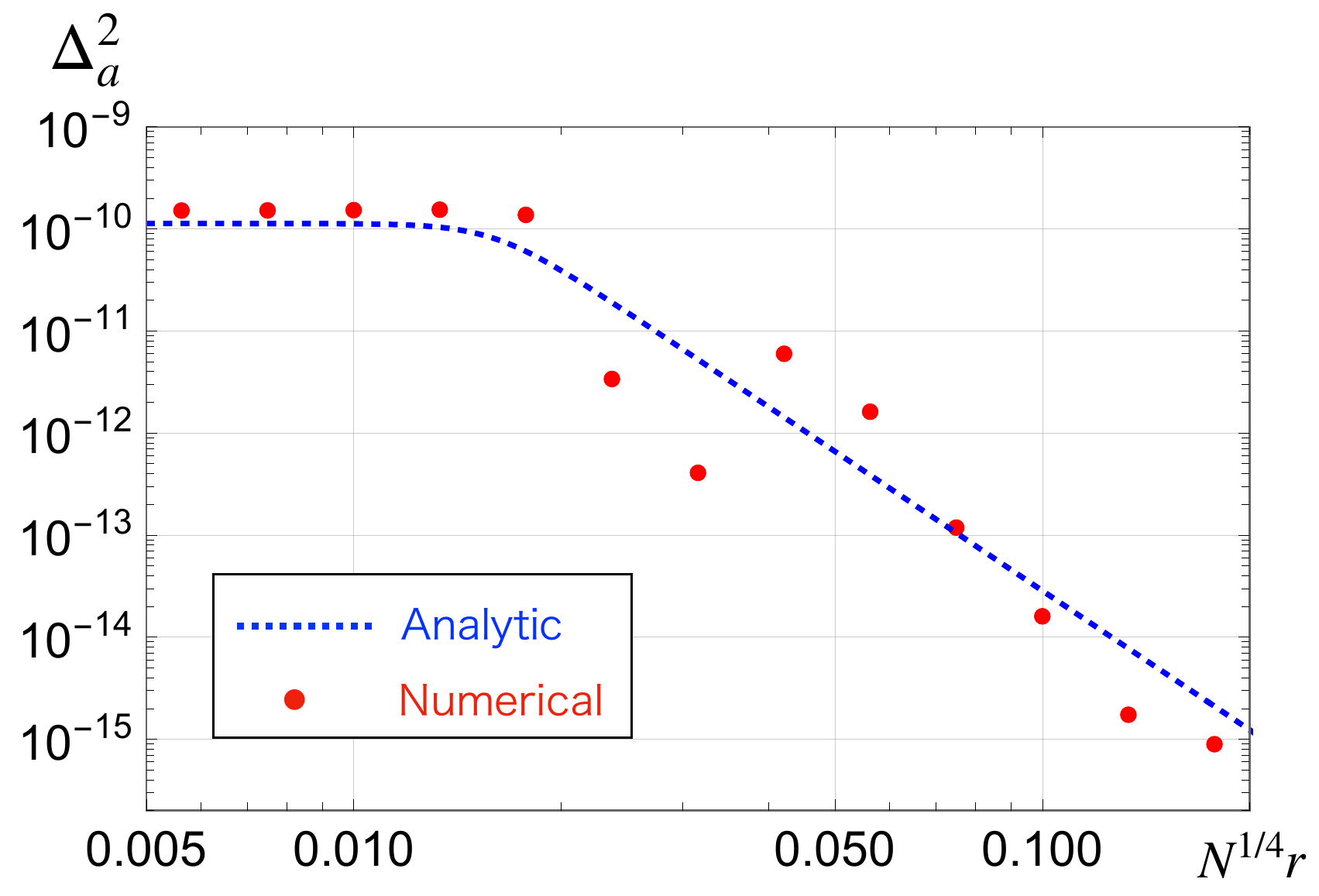}
\centering
\caption{The isocurvature power spectrum as a function of $r$ for $f_a=10^{12}\GeV$, $N=3$, $\theta_H=0$, $\theta_{\rm ini}=3/2$, and $H_{\rm inf}=5\times10^{7}\GeV$. The red bullets denote the numerical results. The blue dotted line denotes the analytic formula (\ref{analyticiso2}) with the quadratic approximation.}
\label{fig:result23}
\end{figure}

In the trapping regime, the isocurvature fluctuations are indeed suppressed as expected by our analytical solution, but the dependence on the initial value and the decay constant cannot be reproduced accurately. 
Next, let us check the dependence of on $r$.
In \FIG{fig:result23} we show the $r$-dependence of $\Delta_a^2$ for the trapping regime, where we take $N=3$, $\theta_H=0$, $\theta_{\rm ini}=3/2$, $f_a=10^{12}\GeV$, and $H_{\rm inf}=5\times10^{7}\GeV$. 
The adopted parameters are same as in Fig.~\ref{fig:result1},
from which one can see that the analytical and numerical results agree with each other at $\theta_{\rm ini}=3/2$.
This value of $\theta_{\rm ini}$ was chosen in order to check numerically the $r$-dependence of the analytical solution of the isocurvature fluctuation.
The red bullets denote the numerical results, and the blue dotted line denotes the analytic solution (\ref{analyticiso2}) with the quadratic approximation. At $r\gtrsim0.01$ where (\ref{interest}) is satisfied, one can see that the isocurvature perturbation is more suppressed for higher $r$. This is because the axion is trapped at a wrong vacuum for a longer time for higher $r$, i.e. $T_{\rm osc}\gg T_{\rm osc2}$, and the axionic fluctuations are suppressed due to the cosmic expansion. 
Thus, the dependence of the isocurvature perturbation on $r$ is well explained by the analytical estimate. 
But we note that the overall agreement is partly due to our choice of $\theta_{\rm ini}$, as described above.

\begin{figure}[t!]
\includegraphics[width=14cm]{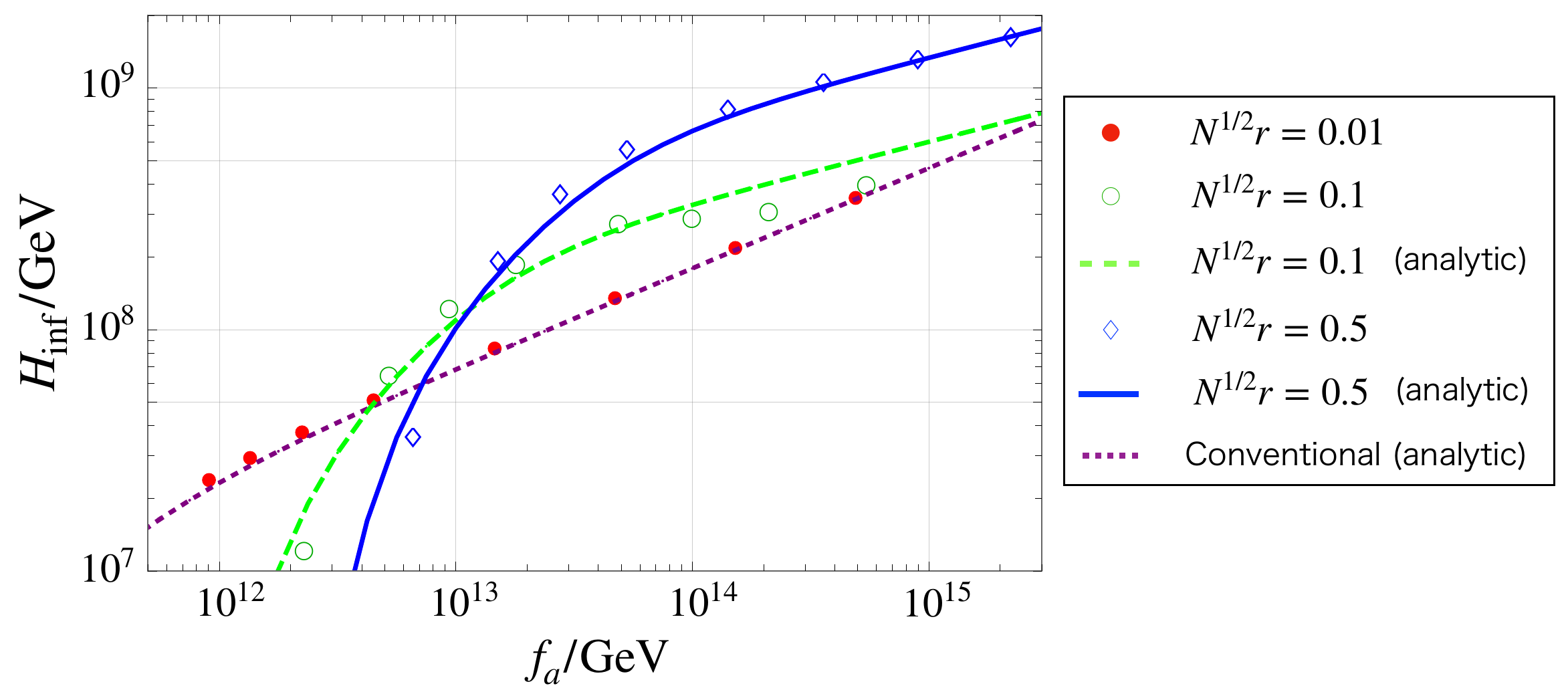}
\centering
\caption{The upper bound on the Hubble parameter $H_{\rm inf}$ during inflation as a function of $f_a$ for the smooth shift regime. We take $N=3$, $\theta_H=0$, and $\Omega_a=\Omega_{\rm DM}$. The red bullet, green circle, and blue diamond denote the numerical results for $N^{1/2}r=0.01, 0.1,$ and $0.5$. The green dashed and blue solid line denote the analytical results for $N^{1/2}r=0.1$ and $0.5$, respectively. The purple dotted line represents the analytical result for the conventional QCD axion.}
\label{fig:faH1}
\end{figure}

Lastly let us derive the isocurvature bound on the axion DM
in our scenario. Using the observational bound (\ref{isoconstraint}), we obtain \FIG{fig:faH1}, which shows the upper bound on the Hubble parameter $H_{\rm inf}$ during inflation as a function of $f_a$ in the case of the smooth shift regime. We set $N=3$, $\theta_H=0$, and $\Omega_a=\Omega_{\rm DM}$. The last condition fixes  $\theta_{\rm ini}$. The red bullet ($\bullet$), green circle ($\circ$), and blue diamond ($\diamondsuit$) denote the numerical results for $N^{1/2}r=0.01, 0.1,$ and $0.5$, respectively. In each case, numerical results correspond to the result for $\theta_{\rm ini}=1, 0.8, 0.6, 0.4, 0.2, 0.1, 0.05, 0.025$ in the order from the left point.
The green dashed and blue solid line denote the analytical results for $N^{1/2}r=0.1$ and $0.5$, respectively. The purple dotted line represents the analytical result for the conventional QCD axion. Here we have taken into account the anharmonic effect.
One can see that the isocurvature bound on $H_{\rm inf}$ is relaxed compared to the conventional case, and we have $H_{\rm inf}\lesssim10^{8}-10^{9}\GeV$ for $f_a\gtrsim10^{13}\GeV$. This is because the axion abundance is suppressed due to the adiabatic suppression mechanism and the early oscillations, and $\theta_{\rm ini}$ increases in order for the axion abundance to explain DM. Note that $\Delta_S^2 \propto 1/\theta_{\rm ini}^2$ in the harmonic approximation. For $f_a \lesssim 10^{12}\GeV$, the isocurvature bound becomes tighter than the usual case because of the anharmonic effect, because we need to put the initial position of the axion near top of the PQ breaking term to explain DM.

\begin{figure}[t!]
\includegraphics[width=14cm]{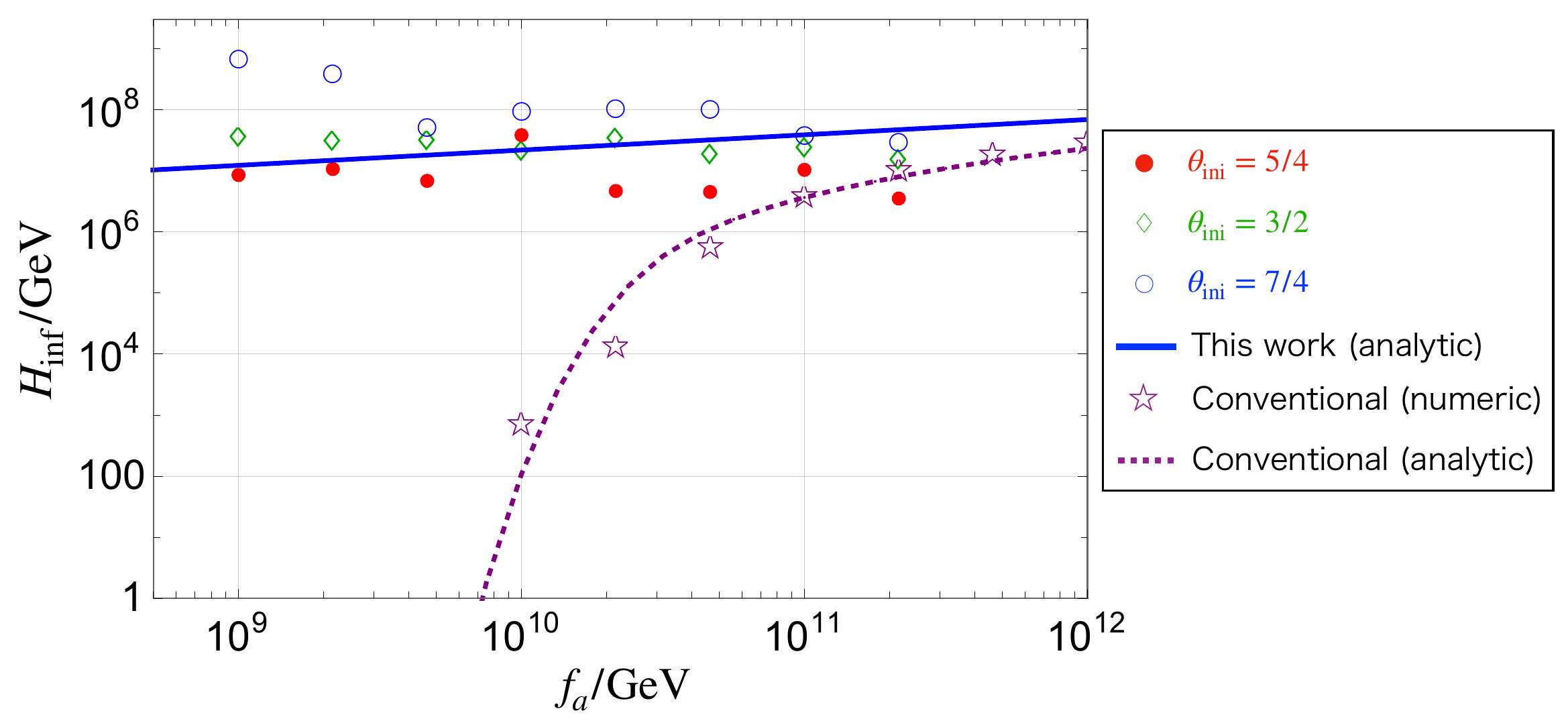}
\centering
\caption{The upper bound on the Hubble parameter $H_{\rm inf}$ during inflation as a function of $f_a$ for the trapping regime. We take $N=3$, $\theta_H=0$, and $\Omega_a=\Omega_{\rm DM}$. The red bullet ($\bullet$), green diamond ($\diamondsuit$), and blue circle ($\circ$) denote the numerical results for $\theta_{\rm ini}=5/4, 3/2$, and $7/4$, respectively. The blue solid line represents the analytical result (\ref{analyticiso2}). The purple star ($\star$) and the purple dotted line denote the numerical and analytical results for the conventional case, respectively. The gray shaded region is the lower bound on $f_a$ from astrophysical facts.
}
\label{fig:faH2}
\end{figure}

We show in \FIG{fig:faH2} the isocurvature bound on $H_{\rm inf}$ for the trapping regime.  We take $N=3$, $\theta_H=0$, and $\Omega_a=\Omega_{\rm DM}$ which sets $N^{1/4}r\sim0.02$. The red bullet ($\bullet$), green diamond ($\diamondsuit$), and blue circle ($\circ$) denote the numerical results for $\theta_{\rm ini}=5/4, 3/2$, and $7/4$, respectively. The blue solid line represents the analytical result (\ref{analyticiso2}). The purple star ($\star$) and the purple dotted line denote the numerical and analytical results for the conventional case, respectively. 
Interestingly, the isocurvature bound is much more relaxed than the conventional result especially for small $f_a \lesssim {\cal O}(10^{10})\GeV$. This is because of the suppression of the axionic fluctuation by the cosmic expansion. As expected, the analytical results agree reasonably well with the numerical ones, but do not fully explain the dependence on $f_a$ and $\theta_{\rm ini}$.

\section{Discussion and conclusions
\label{sec:conclusion}}

We have thus far focused on the case of $N=3$ in our numerical calculations. 
In the case of $N=2$, when the axion is trapped at a wrong vacuum $\theta=\pi$, the final abundance can be strongly enhanced due to the anharmonic effect in addition to the trapping effect, because the minimum turns into a potential maximum at $T\lesssim T_{\rm osc2}$. In the case of even integer $N$, the situation is
similar, and one of the vacua is located near $\theta = \pi$, if $|\theta_H|\ll 1$.
A similar situation was studied in Ref.~\cite{Kitajima:2020kig} using the Witten effect,  where the enhanced abundance of the axion produces primordial black holes.
On the other hand, the usual enhancement of the isocurvature perturbations due to the anharmonic effect is expected to be milder because of the early oscillations.
These issues will be studied elsewhere.

Let us briefly comment on the high quality problem of the PQ symmetry.
Our work has opened up a new parameter region allowed for the axion DM. In particular, the axion abundance can be significantly enhanced in the trapping regime due to the extra PQ breaking, and its abundance is independent of $f_a$. 
As a result, if we assume the anthropic bound on the axion DM is close to the observed DM abundance,
the size of the PQ breaking should satisfy $r \lesssim 0.02$. Thus,
the quality problem is equivalent to fine-tuning the relative phase as $|\theta_{\rm H}| \lesssim 10^{-3}$ (see Fig.~\ref{fig:contour2}).
One of the important implications we have obtained is that the quality problem of the PQ symmetry
becomes closely related to the anthropic argument on the axion DM abundance when there is extra breaking of the PQ symmetry.

Throughout this paper we have focused on the pre-inflationary scenario where the PQ symmetry is already spontaneously broken during inflation. One can also consider the post-inflationary scenario where the PQ symmetry is spontaneously broken after inflation. In this case there appear the axionic strings as topological defects at the phase transition. In  contrast to the ordinary scenario, domain walls separating the different vacua of $V_{\cancel{{\rm PQ}}}$ appear when the axion starts to oscillate, if $T_{\rm osc} > T^{\rm (conv)}_{\rm osc}$. Then the domain walls experience the energy bias and some of them disappear when $V_{\rm QCD}$ becomes important. If the domain wall number is equal to unity, the entire string-wall network decay into axions. The decay process of the string-wall network is analogous to that considered in Ref.~\cite{Sato:2018nqy}, where the domain walls appear due to the extra PQ breaking term induced by hidden monopoles via the Witten effect~\cite{Witten:1979ey}. The evolution of the string-wall network is thus significantly affected by the PQ breaking term, and it is worth studying its effect on the abundance of axions produced from those topological defects.

The parameter region for DM axion is extended by the explicit PQ breaking effect. In the trapping regime, the DM abundance can be explained independently of $f_a$ if $r\sim0.02$. In particular, it is very attractive that the axion with $f_a\lesssim10^{11}\GeV$ is allowed.  Such a relatively heavy axion coupled to photon can be searched for by DM axion search experiments, such as ADMX \cite{Stern:2016bbw}, MADMAX \cite{Caldwell:2016dcw}, ORGAN \cite{McAllister:2017lkb}, and TOORAD \cite{Marsh:2018dlj}. 
Moreover, a new way to detect DM axion with $f_a\sim10^{10}-10^{11}\GeV$ using a correlation with ``condensed matter axion" has been proposed recently \cite{Chigusa:2021mci}. We have also shown that, in the smooth shift regime, the axion abundance is suppressed, and so one can relax the fine-tuning of $\theta_{\rm ini}$ to have the right DM abundance by axions for $f_a\gtrsim10^{13}\GeV$. Such relatively light axions can be searched for by ABRACADABRA \cite{ABRACADABRA:2018rtf}
or DMRadio \cite{DMradio,DMradio2}, and KLASH \cite{Alesini:2017ifp}.

In this paper we have studied cosmological effects of explicit PQ breaking on the QCD axion DM. We have opened up a new parameter region. In the smooth shift regime, $|\theta_{\rm ini}-\theta_H|\lesssim\pi/N$, the axion with $f_a\gtrsim10^{12}\GeV$ can explain dark matter with a milder tuning of $\theta_{\rm ini}$ than the conventional case, thanks to the adiabatic suppression. In the trapping regime, $|\theta_{\rm ini}-\theta_H|\gtrsim\pi/N$, if $r\simeq0.02$ and $f_a\lesssim10^{11}\GeV$, the axion explains all DM, almost independent of $f_a$. Both cases have important implications for the current or projected axion search experiments.
We have also estimated the isocurvature power spectrum analytically and numerically. In both regimes, the isocurvature perturbation can be  suppressed compared to the conventional QCD axion without PQ breaking potentials. The upper bound on $H_{\rm inf}$ is given by $\sim10^{8}-10^9\GeV$ in the smooth shift regime and $\sim10^7-10^8\GeV$ in the trapping regime.
We emphasize that in the trapping regime, since the isocurvature perturbation is  suppressed without the enhancement due to the anharmonic effect, the isocurvature bound can be relaxed by many orders of magnitude with respect to the normal scenario. Together with the fact that
the final axion abundance is independent of $f_a$, we have opened up a scenario that allows the axion with $f_a\lesssim10^{11}\GeV$
to explain all DM without running afoul of the isocurvature bounds. The price we have to pay is to introduce an extra PQ breaking term
with $N^{1/4}r \approx 0.02$ and $|\theta_H| \lesssim 10^{-3}$. This requires a mild tuning of the relative phase of the PQ breaking. Conversely, if $\theta_H$ is near the upper limit, it could be observed through nEDM in the near future.

\section*{Acknowledgments}
The present work is supported by 
the Graduate Program on Physics for the Universe of Tohoku University (S.N.), JST SPRING, Grant Number JPMJSP2114 (S.N.),  Leading Young Researcher Overseas Visit Program at Tohoku University (F.T.),
JSPS KAKENHI Grant Numbers
17H02878 (F.T.), 20H01894 (F.T.) and 20H05851 (F.T.),
and also by the National Research Foundation (NRF) of
Korea grant funded by the Korea government: Grants
No. 2018R1C1B6006061 (K.S.J.) and No. 2021R1A4A5031460
(K.S.J.).

\appendix

\section{Useful equations for numerical calculations
\label{app:eom}}
Here we derive a useful equation of motion for numerical calculations. First let us rewrite the equation of motion (\ref{eom}) using the dimensionless time $\tau\equiv T_n/T$, where $T_n$ is an arbitrary normalization factor of mass-dimension one \cite{Nakagawa:2020zjr}. Taking account of the temperature-dependence of $g_*$ and $g_{*s}$, we obtain the time derivative of the temperature
\beq
\frac{dT}{dt}&=&-\frac{\pi}{\Mpl}\sqrt{\frac{g_*(T)}{10}}\frac{s(T)}{s'(T)}T^2\nonumber\\
&=&-HT\left(\frac{3\tau K(\tau)}{\sqrt{g_*(\tau)}}\right),
\eeq
where $K(\tau)$ is defined as
\beq
K(\tau)\equiv\frac{\sqrt{g_*(\tau)}}{3\tau}\frac{g_{*s}(\tau)}{g_{*s}(\tau)-\tau g'_{*s}(\tau)/3}.
\eeq
We refer to \REF{Saikawa:2018rcs} for the detailed temperature dependence of $g_*$ and $g_{*s}$.
Assuming $g_{*s}$ does not depend on the temperature, one can see that this relation is consistent with the usual one, $dT/dt=-HT$ or $H=1/2t$. In other words, $K(\tau)$ represents the time-dependence of $g_{*s}$. Using this relation, we can obtain the derivative of the time $t$
\beq
\frac{d}{dt}=\sqrt{\frac{\pi^2T_n^4}{10\Mpl^2}}K(\tau)\frac{d}{d\tau}.
\eeq
Thus the equation of motion (\ref{eom}) becomes
\beq
K^2(\tau)\frac{d^2\theta}{d\tau^2}&+&K(\tau)\left[\frac{dK}{d\tau}+\sqrt{g_*}\tau^{-2}\right]\frac{d\theta}{d\tau}+\frac{10\Mpl^2}{\pi^2T_n^4f_a}V'(a)=0,
\label{generaleom}
\eeq
where $\theta\equiv a/f_a$. This equation is applicable to a homogeneous scalar field with any differentiable potential. Note that in the above derivation we use only the fact $sR^3(T)={\rm const.}$ with $R(T)$ a scale factor, so (\ref{generaleom}) generally holds. We substitute the potential (\ref{potential}) for (\ref{generaleom}) and obtain the equation
\beq
K^2(\tau)\frac{d^2\theta}{d\tau^2}&+&K(\tau)\left[\frac{dK}{d\tau}+\sqrt{g_*(\tau)}\tau^{-2}\right]\frac{d\theta}{d\tau}\nonumber\\
&+&\frac{10\Mpl^2}{\pi^2T_n^4}\left[m_a^2(\tau)\sin\theta+Nr^4m_{a,0}^2\sin\left(N\left(\theta-\theta_H\right)\right)\right]=0.
\label{oureom}
\eeq

\bibliography{reference}

\end{document}